\documentclass[11pt,english]{article}
\usepackage{float}
\usepackage{calc}
\usepackage{textcomp}
\usepackage{amsmath}
\usepackage{graphicx}
\usepackage{setspace}
\usepackage[authoryear]{natbib}
\usepackage{pdflscape}
\makeatletter
\usepackage{caption}
\usepackage{setspace}
\usepackage{rotating}
\usepackage{threeparttable}

\usepackage{booktabs}
\usepackage{multicol}
\usepackage{multirow}
\usepackage{dcolumn}
\newcolumntype{.}{D{.}{.}{-1}}

\usepackage{tikz}
\usetikzlibrary{patterns}
\usetikzlibrary{patterns.meta}

\usepackage{lscape}

\usepackage{hyperref}
\hypersetup{colorlinks=true,linkcolor=darkblue,urlcolor=darkblue,citecolor=darkblue,urlbordercolor={1 1 1}}
\usepackage{amsmath,amsfonts,bbm,amsthm}

\makeatother
\usepackage{babel}
\usepackage{pdfpages}

\usepackage{appendix}

\usepackage{array}
\usepackage{booktabs}

\usepackage[T1]{fontenc}%
\usepackage{times}

\usepackage{tikz}
\usepackage{authblk}
\usepackage{indentfirst}

\usepackage{amsthm}
\newtheorem{theorem}{Theorem}[section]

\newtheorem{remark}[theorem]{Remark}

\newcommand{\MX}{\mathcal{X}}
\newcommand{\MW}{\mathcal{W}}

\newcommand{\MG}{\mathcal{G}}

\newcommand{\Real}{\mathbb{R}}

\newcommand{\argmax}{\arg\max}

\tikzstyle{rednode} = [shape=rectangle, fill=red!50, line width=3]
\tikzstyle{bluenode} = [shape=rectangle, fill=blue!50, line width=3]
\tikzstyle{yellownode} = [shape=rectangle, fill=yellow!50, line width=3]
\tikzstyle{dotnode} = [dashed, pattern={Lines[angle=90,distance=3pt]}, pattern color=gray!150]
\tikzstyle{backslashnode} = [dashed, pattern={Lines[angle=45,distance=3pt]}, pattern color=gray!150]
\tikzstyle{slashnode} = [dashed, pattern={Lines[angle=-45,distance=3pt]}, pattern color=gray!150]

\usepackage{geometry}
\begin{document}
\definecolor{darkblue}{rgb}{0.0, 0.0, 0.7}

\begin{spacing}{1.5}

\title{\vspace{-50px} Paternalism, Autonomy, or Both? \\ Experimental Evidence from Energy Saving Programs%
\thanks{Ida and Kido: Graduate School of Economics, Kyoto University, Yoshida, Sakyo, Kyoto 606-8501, Japan (e-mails: ida@econ.kyoto-u.ac.jp and daido.kido@gmail.com). Ishihara: Faculty of Economics and Business Administration, Kyoto University of Advanced Science, 18 Yamanouchi Gotanda, Ukyo, Kyoto 615-8577, Japan (e-mail: ishihara.takunori@kuas.ac.jp). Ito: Harris School of Public Policy, University of Chicago, 1307 East 60th St., Chicago, IL 60637 (e-mail: ito@uchicago.edu). Kitagawa: Department of Economics, Brown University, 64 Waterman St., Providence, RI 02912 (e-mail: toru\_kitagawa@brown.edu). Sakaguchi: Department of Economics, University College London, 30 Gordon Street, London WC1H 0AX, United  Kingdom (e-mail: s.sakaguchi@ucl.ac.uk). Sasaki: Faculty of Economics, Tohoku Gakuin University, 1-3-1 Tsuchitoi, Aoba-ku, Sendai, Miyagi 980-8511, Japan (ssasaki.econ@gmail.com). We would like to thank Severin Borenstein, Fiona Burlig, Mark Jacobsen, Louis Preonas, Frank Wolak, and seminar participants at UC Berkeley for their helpful comments. We thank the Japanese Ministry of Environment for their collaboration for this study. Kitagawa and Sakaguchi gratefully acknowledge financial support from ERC grant (number 715940) and the ESRC Centre for Microdata Methods and Practice (CeMMAP) (grant number RES-589-28-0001). Ito gratefully acknowledge support from Research Institute of Economy, Trade and Industry, and note that this project was conducted as part of a research project ``Empirical Research on Energy and Environmental Economics".%
}}

\author[1]{Takanori Ida}
\author[2]{Takunori Ishihara}
\author[3]{Koichiro Ito}
\author[1]{Daido Kido}
\author[4]{\\  Toru Kitagawa} 
\author[5]{Shosei Sakaguchi}
\author[6]{Shusaku Sasaki}
\affil[1]{\small{Kyoto University}}
\affil[2]{Kyoto University of Advanced Science}
\affil[3]{University of Chicago and NBER}
\affil[4]{Brown University}
\affil[5]{University College London}
\affil[6]{Tohoku Gakuin University}

\date{\vspace{-10px} This version: \today.}

\maketitle
\vspace{-50px}
\end{spacing}
\begin{spacing}{1.1}
\begin{abstract}
\noindent %
Identifying who should be treated is a central question in economics. There are two competing approaches to targeting---\textit{paternalistic} and \textit{autonomous}. In the paternalistic approach, policymakers optimally target the policy given observable individual characteristics. In contrast, the autonomous approach acknowledges that individuals may possess key unobservable information on heterogeneous policy impacts, and allows them to self-select into treatment. In this paper, we propose a new approach that mixes paternalistic assignment and autonomous choice. Our approach uses individual characteristics and empirical welfare maximization to identify who should be treated, untreated, and decide whether to be treated themselves. We apply this method to design a targeting policy for an energy saving programs using data collected in a randomized field experiment. We show that optimally mixing paternalistic assignments and autonomous choice significantly improves the social welfare gain of the policy. Exploiting random variation generated by the field experiment, we develop a method to estimate average treatment effects for each subgroup of individuals who would make the same autonomous treatment choice. Our estimates confirm that the estimated assignment policy optimally allocates individuals to be treated, untreated, or choose themselves based on the relative merits of paternalistic assignments and autonomous choice for individuals types. 
\end{abstract}
\end{spacing}
\thispagestyle{empty}
\newpage{}
\restoregeometry

\setcounter{page}{1}
\doublespacing

\section{Introduction}

Targeting has become a central question in economics and policy design. When policymakers face budget constraints, identifying those who should be treated is critical to maximizing policy impacts. Advances in machine learning and econometric methods have led to a proliferation of targeting methods, which has contributed to a surge in research on policies including job training programs \citep{KT18}, social safety net programs \citep{Finkelstein_2018, Deshpande_2019}, energy efficiency programs \citep*{burlig2020machine}, behavioral nudges for electricity conservation \citep{knittel2019using}, and dynamic electricity pricing \citep*{Ito_2019}.

In this paper, we begin by highlighting the two competing approaches to effective policy design in the literature ---\textit{paternalistic} and \textit{autonomous}. The paternalistic approach involves policymakers using individuals' observable characteristics to  target the program optimally\citep*{KT18, zhou_et_al_2018}. In contrast, the autonomous approach exploits self-selection to accomplish effective targeting under the belief that individuals' own choices take into account unobservable information on heterogeneous policy impacts \citep*{Heckman_2005, Heckman_2010, Ito_2019}. 

A priori, it is ambiguous which approach is superior. For example, referring to the paternalistic and autonomous approaches as ``planner's decisions" and ``laissez-faire", \cite{ManskiBook13} summarizes their relative merits as: 

\begin{quote}
\textit{``The bottom line is that one should be skeptical of broad assertions that individuals are better informed than planners and hence make better decisions. Of course, skepticism of such assertions does not imply that planning is more effective than laissez-faire. Their relative merits depend on the particulars of the choice problem."} \\
\null\hfill ---Charles F. Manski, \textit{Public Policy in an Uncertain World}   
\end{quote}
A common view in the literature, reflected in this quote, is that the appropriate approach depends on the context, and therefore, researchers and policymakers need to decide which to use on a case-by-case basis.  

We propose a new approach that mixes paternalistic assignment and autonomous choice. Consider a treatment where treating an individual is costly but may generate a social welfare gain. That is, the net social welfare gain from assigning a particular individual to treatment can be positive or negative. We consider using observable individual characteristics to identify three types: 1) those who should be treated, 2) those who should not be treated, and 3) those who should decide whether to receive treatment themselves. 

We build an algorithm based on the empirical welfare maximization (EWM) method developed by \cite{KT18}. We show that the optimal targeting policy of our approach can be identified and estimated using data obtained from a randomized controlled trial (RCT) or a quasi-experiment with a double randomization design. If we can obtain data from an RCT in which individuals are randomly assigned to one of the three arms (compulsory treatment, compulsory no-treatment, and opt-in treatment), we can characterize the expected welfare gain for each observable individual type for each arm. With this information in hand, we can identify the optimal targeting policy by maximizing an empirical welfare criterion over a class of assignment policies such as policy trees \citep*{zhou_et_al_2018}.    

We apply this method to data from a randomized field experiment involving an energy conservation program. This experiment was conducted in partnership with the Japanese Ministry of the Environment in the summer of 2020, and involved 3,870 households. Customers participating in the experiment were eligible to receive a rebate by reducing their electricity consumption during peak demand hours (1 pm to 5 pm) on days when the system faced supply shortages relative to demand. Because there is a per-household implementation cost for this policy, the net social welfare gain from treating an individual can be positive or negative. 

We provide several key findings. First, our experimental results indicate that there is substantial heterogeneity in treatment effect over observable household characteristics. As an example, we compare intention-to-treat (ITT) estimates for the compulsory treatment group (100\% of households were treated) and the opt-in group (households self-selected into treatment). For some values of the observable variables, one of these ITTs is larger than the other and the difference between the two is statistically significant. However, for other values of the observable variables, this relationship does not exist or flips. This implies that the relative effectiveness of compulsory and opt-in treatment can vary substantially by observable household type, which suggests that the optimal policy assignment differs across households. 

Building upon this insight, we use our algorithm to identify the optimal assignment for each household type. We begin with a conventional paternalistic approach, where we assign consumers to either of treatment or no-treatment. Our algorithm finds that 52\% of households should be treated and 48\% should be untreated. This optimal policy assignment with two assignment options significantly improves social welfare relative to treating all households or treating no households. 

We then ask if adding an opt-in group (i.e., households who self-select into treatment) further improves welfare. Our algorithm finds that 37\% of households should be treated, 19\% should be untreated, and 44\% should be assigned to the opt-in group. Optimal policy assignment with three assignment options further improves the social welfare relative to the paternalistic approach described above. These results imply that optimally mixing paternalistic and autonomous approaches can maximize the welfare gain from a policy. 

What is the mechanism behind these findings? To explore this question, we develop new methods that exploit the random variation generated by the field experiment. The key insight that our RCT created three randomly-assigned groups (treatment, control, and opt-in) for each observable household types. We can use this variation to estimate two local average treatment effects (LATEs) that provide insight into the mechanism behind our findings. 

The first LATE is the ATE for the \textit{takers} from the opt-in group: households who would take the treatment if assigned to the opt-in group. The second LATE is the ATE for the non-takers from the opt-in group: households who would not take the treatment if assigned to the opt-in group. Using the random variation from the RCT, we can estimate these two LATEs for each observable household types. We then examine how these two LATEs differ between the three groups identified by our algorithm. 

For households our algorithm assigns to the opt-in group, we find that the LATE for takers is positive and large, while the LATE for non-takes is negative. 
That is, for these households, self-selection is very informative about the effectiveness of the treatment. In contrast, for households assigned to the (compulsory) treatment group, both LATEs are positive. In this group, there is a positive welfare gain from treating both takers and non-takers, so self-selection would not improve social welfare. Similarly, for households assigned to (compulsory) no-treatment, both LATEs are negative. For these households too, self-selection would not improve social welfare, and it is optimal for policymakers to leave them untreated. 

Using the same method, it is also possible to calculate the ITTs for three counterfactual outcomes (outcomes if assigned to treatment, control, or opt-in) for each group. We confirm that the optimal groud assignment maximizes these ITTs, and alternative treatment assignments would reduce the ITTs of the welfare gain. 

\textit{Related literature and our contributions}---Our study is related to three strands of the literature. First, many recent studies in the economics literature have explored targeting based on paternalistic or autonomous approaches. In addition to the papers cited earlier, recent studies using paternalistic assignment include \cite*{johnson_2020, murakami_2020, cagala_2021,christensen_2021, gerarden_2021} and studies using autonomous approaches include \cite*{alatas_2016, dynarski_2018, lieber_2019, unrath_2021, waldinger_2021}. We are not aware of any existing study that builds an algorithm to identify the optimal mix of paternalistic assignment and autonomous choice.   

Second, the medical statistics literature has studied hybrid sampling designs that combine randomization and treatment choice by patients. See, e.g., \cite*{Janevic_2003}, \cite*{Long_etal_2008}, and references therein. In this literature, the sampling process used in our experiment is referred to as a doubly randomized preference trial \citep*{Rucker_1989}. An example of a clinical trial that implements a doubly randomized preference design is the Woman Take Pride study analyzed in \cite{Janevic_2003}. These studies focus on assessing whether letting patients choose their own treatment can have a direct causal effect on their health status beyond the causal effect of the treatment itself. See \cite*{Knox_etal_2019} for partial identification analysis in such a context and an application to political science. Double randomized preference trials have received less attention in economics. The only paper we are aware of is \cite{Bhattacharya_2013}, which uses double randomization between randomized control trials and planner's allocation to assess the efficiency of the planner's treatment allocations. To our knowledge, no work has analyzed double randomized preference trial data to estimate an optimal targeting policy which mixes paternalistic assignment and autonomous choice.   

Third, the econometric framework we use in our empirical analysis builds on the growing  statistical treatment choice literature. Generally assuming discrete characteristics, earlier studies in this literature \citep[][among others]{manski2004statistical,Dehejia2005,HiranoPorter2009,stoye2009minimax, stoye2012minimax,Chamberlain2011,tetenov2012statistical} formulate estimation of a treatment assignment rule as a statistical decision problem. The empirical welfare maximization approach proposed by \cite{KT18} estimates a treatment assignment rule by maximizing the in-sample empirical welfare criterion over a class of assignment rules. This approach can accommodate multi-armed treatment assignment and a rich set of household characteristics, including continuous characteristics, as is the case for our empirical application. We employ a class of tree partitions considered in \cite{AW20} and \cite{zhou_et_al_2018} as our class of policy rules. 

\section{Conceptual Framework}

Section \ref{sec:optimal_arm_assignment} formulates an optimal arm assignment problem that allows for a mixture of paternalism and autonomous choice, and characterizes the optimal arm assignment given observed individual characteristics. Section \ref{sec:EWM_tree} describes how the EWM method with decision trees can be applied to estimate the optimal arm assignment from RCT data. 

\subsection{Optimal Arm Assignment: Paternalism, Autonomy, or Both?}\label{sec:optimal_arm_assignment}

Consider a planner who wishes to introduce a policy intervention (program) to a population of interest. Instead of the uniform assignment over the entire population, the planner is interested in targeted assignment over heterogeneous individuals. A novel feature of our setting is that the planner can control not only who is compulsorily exposed to the program but also who is given an option to opt-in to the program. Interpreting an individual's take-up of the program as her exposure to the treatment, the planner's goal is therefore to assign each individual in the population to one of the three arms: \textit{compulsory treatment} (indexed as $T$), \textit{compulsory no-treatment} (indexed as $NT$), and \textit{opt-in treatment} (indexed as $O$). An individual assigned to compulsory treatment or no-treatment is exposed to or excluded from the program with no opt-out or opt-in option, whereas an individual assigned to opt-in treatment chooses whether to take it up by herself. 

The planner's goal is to optimize a social welfare criterion by assigning individuals to these three arms. Following the statistical treatment choice literature \citep{manski2004statistical}, we specify the planner's social welfare criterion to be the sum of individuals' welfare contributions. An individual's welfare contribution is a known function of the individual's response to being assigned to arm $T$, $NT$, or $O$, and the per-person cost of the treatment. An individual's welfare contribution may not correspond to her utility. Hence, if an individual is assigned to opt-in treatment, her utility maximizing decision may not correspond to the choice that maximizes the planner's objective.

Let $W(T)$, $W(NT)$, and $W(O)$ denote the potential welfare contributions that would be realized if an individual were assigned to compulsory treatment, compulsory no-treatment, and opt-in treatment.

We assume that the planner observes a pre-treatment characteristic vector for each individual $x \in \MX$, where $\MX$ denotes the support of the characteristics. Depending on these observable characteristics, the planner assigns each individual to one of the three arms. Let $G_{T} \subseteq \MX$ denote a set of the pre-treatment characteristics $x$ such that any individual whose $x$ belongs to $G_T$ is assigned to the compulsory treatment.
Similarly, let $G_{NT}$ and $G_{O}$ denote sets of the pre-treatment characteristics $x$ such that the individuals with $x \in G_{NT}$ are assigned to compulsory no-treatment and individuals with $x \in G_O$ are assigned to opt-in treatment, respectively. We consider non-randomized assignment policies only, so $G_{T}$, $G_{NT}$, and $G_{O}$ can be interpreted as a partition of $\MX$, i.e.,  $G_{T}$, $G_{NT}$, and $G_{O}$ are disjoint and $G_{T}\cup G_{NT} \cup G_{O} = \MX$ (see, e.g., Figures \ref{fig:example_decision_trees} (a-2) and (b-2)).

We call a partition $G:=(G_{T},G_{NT},G_{O})$ an assignment policy. $G$ describes how individuals are assigned to arms according to their observable characteristics $x$. 
The realized welfare contribution after assignment for an individual with characteristics $x$ is either $W(T)$, $W(NT)$, or $W(O)$ depending on $x \in G_{T}$, $x \in G_{NT}$, or $x \in G_{O}$. Hence, her welfare contribution under the policy $G$ can be written as 
\begin{equation}
\sum_{j \in \{T,NT,O\}}W(j) \cdot 1\{x \in G_{j}\}.
\end{equation}
Viewing individual characteristics and their potential welfare contributions as random variables, the average welfare contribution under assignment policy $G$ can be written as
\begin{align*}
    \mathcal{W}(G) \equiv E\left[\sum_{j \in \{T,NT,O\}}W(j) \cdot 1\{X \in G_{j}\} \right],
\end{align*}
where the expectation is with respect to $(W(T),W(NT),W(O),X)$.
We define $\mathcal{W}(G)$ as our social welfare function. The social welfare function depends on the assignment policy $G$ through the post-assignment distribution of individual welfare contributions, which can be manipulated by changing which individuals are assigned to which arms. This form of social welfare is standard in the statistical treatment choice literature. Note that, since we allow the individual welfare contributions to differ from individual utilities, the interpretation of social welfare is not restricted to be utilitarian. The planner's objective is to find an assignment policy $G^{\ast}$ that maximizes the social welfare $\mathcal{W}(G)$ over a set of possible assignment policies. If the planner can implement any assignment policy, this set of assignment policies corresponds to the set of measurable partitions of $\MX$. Accordingly, $G^{\ast}$ can be defined by
\begin{align}
    G^{\ast} \in \argmax_{G \in \widetilde{\MG}}\mathcal{W}(G), \label{eq:optimal_policy}
\end{align}
where $\widetilde{\MG} := \{G=(G_{T},G_{N},G_{O}):  G \mbox{ is a measurable partition of }\MX\}$.

It is desirable that individuals with characteristics $x$ be assigned to an arm that gives the largest conditional mean welfare contribution among $\{E[W(j)\mid x] : j \in \{T,NT,O\}\}$. In the absence of an opt-in treatment arm, the planner's paternalistic assignment policy is to allocate them to either compulsory treatment $T$ or compulsory no-treatment $NT$. 
The optimal choice is then determined by comparing $E[W(T)\mid x]$ and  $ E[W(NT)\mid x]$. That is, an optimal \textit{paternalistic} assignment policy exploits only heterogeneity in the average welfare contribution conditional on observable characteristics $x$, which can be assessed by the planner prior to assignment.

Once individuals are permitted to self-select into treatment, social welfare can be improved beyond the level attained by paternalistic assignment. This is because an individual may possess private information, which drives or helps predict her response to the treatment, and choose whether to receive treatment based on it. On the other hand, such private information is unobservable to the planner, and cannot be exploited through paternalistic assignment. If an individual's treatment choice is aligned with the ordering of their welfare contributions, then letting each individual \textit{autonomously} choose between treatment and no-treatment can dominate paternalistic assignment.
This argument, however, relies crucially on the assumption that individuals' objective functions are aligned with their social welfare contributions. If this assumption does not hold, autonomous treatment choice can reduce social welfare, in which case paternalistic assignment may dominate.

Which of paternalistic and autonomous policies performs better in reality? This is an important empirical question whose answer may well vary across contexts. Furthermore, if the planner can differentiate between paternalistic and autonomous intervention across individuals, it is desirable to \textit{mix paternalistic and autonomous intervention} and assign individuals to the three arms, $T$, $NT$, or $O$, according to their observable characteristics $x$. 
It may be that, for some values of $x$, individuals choose by themselves the treatment that is optimal in terms of the social welfare, so that their choices outperform a planner's paternalistic assignment based only on $x$. Assigning such individuals to autonomous opt-in treatment is optimal. In contrast, for values of $x$ where individuals are likely to choose treatment that is suboptimal in terms of the social welfare, autonomous choice will underperform a planner's assignment based on $x$. Paternalistic assignment is then optimal.
Thus, an optimal assignment policy $G^{\ast}$ that mixes paternalistic assignment (i.e., compulsory treatment and no-treatment) and autonomous choice (i.e., opt-in treatment) depending on individual observed characteristics $x$ can improve social welfare compared with paternalistic assignment or autonomous choice alone.

In the following remark, we present a simple model that clarifies how the optimal assignment policy $G^{\ast}$ mixes paternalistic assignment and autonomous choice in accordance with individual observable characteristics $x$.

\begin{remark}\label{rem:mechanism}
Let $Z(O) \in \{T,NT\}$ denote the individual's autonomous choice of treatment. That is, we observe $Z(O)$ if she is assigned to opt-in treatment. The autonomous choice $Z(O)$ may depend on both observable characteristics $X$ and unobservable characteristics (i.e., private information).  We assume that her welfare contribution at the opt-in arm satisfies
\begin{align*}
    W(O) = W(T)\cdot 1\{Z(O)=T\} + W(NT) \cdot 1\{Z(O)=NT\}.
\end{align*}
An implicit assumption here is that an individual's response to the treatment is the same irrespective of whether she opts-in herself or is assigned to it by the planner. 
This is similar to the exclusion restriction for instrumental variables, with an indicator for assignment to the opt-in treatment corresponding to an instrumental variable. We assume this in the current remark only to simplify exposition, and stress that the validity of our method and the estimated optimal assignment policy do not rely on this exclusion restriction.

The optimal arm for individuals with characteristics $x$ attains the highest conditional average welfare contribution among $E[W(T)\mid x]$, $E[W(NT)\mid x]$, and $E[W(O)\mid x]$. Considering paternalistic assignment policies only, compulsory treatment is superior to compulsory no-treatment if and only if 
\begin{align}
    E[W(T) - W(NT)\mid x] > 0. \label{eq:T_vs_NT}
\end{align}
If opt-in treatment is also available, we have the following identities: for $j \in \{T,NT\}$
\begin{align*}
    E[W(j)\mid x] &=E[W(j)\mid Z(O)=T, x] P(Z(O)=T\mid x) + E[W(j)\mid Z(O)=NT, x] P(Z(O)=NT\mid x), \\
    E[W(O)\mid x] &=E[W(T)\mid Z(O)=T, x] P(Z(O)=T\mid x) + E[W(NT)\mid Z(O)=NT, x] P(Z(O)=NT\mid x). 
\end{align*}
These imply the following equivalence relations: 
\begin{align}
    E[W(O)\mid x] \geq E[W(T)\mid x] &\Leftrightarrow E[W(T) - W(NT)\mid Z(O)=NT,x] \leq 0 \mbox{ or } P(Z(O)=T \mid x)=1, \label{eq:O_vs_T} \\
    E[W(O)\mid x] \geq E[W(NT)\mid x] &\Leftrightarrow E[W(T) - W(NT)\mid Z(O)=T,x] \geq 0 \mbox{ or } P(Z(O)=T \mid x)=0. \label{eq:O_vs_NT}
\end{align}
That is, given $\Pr(Z(O)=T|x) \in (0,1)$, opt-in treatment dominates compulsory treatment (no-treatment) if and only if individuals are on average voluntarily making the welfare improving choice, i.e. the mean of $W(T)$ is higher than the mean of $W(NT)$ for those who choose $Z(O)=T$, and the opposite is true for those who choose $Z(O)=NT$. One extreme case where it is immediate that both (\ref{eq:O_vs_T}) and (\ref{eq:O_vs_NT}) hold is a Roy model where an individual's latent utility coincides with their social welfare contribution,
\begin{equation*}
Z(O)=\begin{cases} T & \text{if $W(T) \geq W(NT)$,} \\ NT & \text{if $W(NT) > W(T)$.} \end{cases}    
\end{equation*}
Under this selection equation, $W(O) = \max\{W(T),W(NT)\}$ holds, and both (\ref{eq:O_vs_T}) and (\ref{eq:O_vs_NT}) follow.

Note that an individual's autonomous choices $Z(O)$ and their potential welfare contributions $(W(T),W(NT))$ can depend on private information. 
Thus the three conditional average treatment effects $E[W(T)-W(NT) \mid x]$, $E[W(T)-W(NT) \mid Z(O)=T,x]$, and $E[W(T)-W(NT) \mid Z(O)=NT,x]$ 
in (\ref{eq:T_vs_NT})--(\ref{eq:O_vs_NT}) vary depending the extent to which an individual's treatment choice is correlated with their underlying potential welfare contributions.




By (\ref{eq:T_vs_NT})--(\ref{eq:O_vs_NT}), the optimal arm for individuals with $X=x$ can be determined by comparing three conditional average treatment effects: $E[W(T) - W(NT)\mid x]$, $E[W(T) - W(NT)\mid Z(O)=T, x]$, and $E[W(T) - W(NT)\mid Z(O)=NT, x]$, all of which can vary with observed characteristics $x$.
Specifically, an optimal assignment policy  $G^{\ast}$ as defined in (\ref{eq:optimal_policy}) has the form $G^{\ast} = (G_{T}^{\ast},G_{NT}^{\ast},G_{O}^{\ast})$ with
\begin{align*}
    G_{T}^{\ast} &= \{x \in \MX: E[W(T) - W(NT)\mid x] \geq 0 \mbox{ and } E[W(T) - W(NT)\mid Z(O)=NT, x] > 0\},\\
    G_{NT}^{\ast} &= \{x \in \MX: E[W(T) - W(NT)\mid x] < 0 \mbox{ and } E[W(T) - W(NT)\mid Z(O)=T, x]< 0\},\\
    G_{O}^{\ast} &= \{x \in \MX: E[W(T) - W(NT)\mid Z(O)=T, x] \leq 0 \mbox{ and } E[W(T) - W(NT)\mid Z(O)=NT, x] \geq 0\}.
\end{align*}
Figure \ref{fig:example_optimal_policy} illustrates an example of an optimal policy $G^{\ast}$ for $\MX \subseteq \Real^2$.
\end{remark}

\begin{center}
[Figure \ref{fig:example_optimal_policy} about here]
\end{center}


\subsection{Policy Learning by the EWM with Decision Trees}
\label{sec:EWM_tree}

Our goal is to find an optimal assignment policy $G^{\ast}$ by choosing the optimal arm for each value of observed characteristics $x$. Using data from an RCT in which treatment arms $T$, $NT$, and $O$ are randomly assigned, we apply an EWM method \citep{KT18} to learn an optimal policy $G^{\ast}$. 

Let the RCT data be a size $n$ random sample of $(W_{i},D_{i},X_{i})$, where $D_{i} \in \{T,NT,O\}$ is individual $i$'s (randomly assigned) treatment arm, $W_{i}$ is their observed outcome (welfare contribution), and $X_{i}$ their observable pre-treatment characteristics.
Letting $\{W_{i}(T),W_{i}(NT),W_{i}(O)\}$ denote potential outcomes for individual $i$, the observed outcome $W_{i}$ is subject to $W_{i} = \sum_{j \in \{T,NT,O\}}W_{i}(j)1\{D_{i}=j\}$.
We assume that $\{W_{i}(T),W_{i}(NT),W_{i}(O),X_{i}\}_{i=1,\ldots,n}$ are independently and identically distributed as $\{W(T),W(NT),W(O),X\}$.

Using the RCT data and a class $\MG$ of policies $G$, the EWM method estimates an optimal policy $G^{\ast}$ by maximizing the empirical analogue of the social welfare function over $\MG$:
 \begin{align*}
    \hat{G}_{EWM} &\in \argmax_{G \in \MG } \widehat{\mathcal{W}}(G),\\
    \widehat{\mathcal{W}}(G) &\equiv \frac{1}{n} \sum\limits_{i=1}^{n} \sum_{j \in \{T, C, O\}}  \left(\frac{W_i \cdot 1\{D_i=j\}}{P(D_i=j \mid X_i)} \cdot 1\{X_i \in G_{j}\}  \right),
\end{align*}
where $\widehat{\mathcal{W}}(G)$ is an empirical welfare function of $G$ that produces an unbiased estimate of the population social welfare $\mathcal{W}(G)$. Observations are weighted by the inverse of the propensity scores, $P(D_i=j \mid X_i)$, which are known from the RCT design. 

The EWM approach is a model-free: it does not require any assumptions or a functional form specification for the potential outcome distributions. 
However, the class of policies $\mathcal{G}$ must be specified, taking into account any feasibility constraints for assignment policies. If the class $\MG$ is too rich, the EWM solution $\hat{G}_{EWM}$ will overfit the RCT data, and the social welfare attained by the estimated policy falls.

In the paper, we use a class of decision trees \citep{breiman_et_al2017} as $\MG$. The main reasons for this choice are the ease of interpretation of the decision tree based assignment policies and the availability of partition search algorithms from the classification tree literature. 

To illustrate the interpretation of a decision tree based assignment policy, Figure \ref{fig:example_decision_trees} shows example decision trees of depths 1 and 2 for a two-dimensional $\MX$. By traversing a tree from its top node to a bottom node, we map from $x$ to one of the tree assignment options. This tree structure generates a partition of the characteristic space $\MX$ as in Figures \ref{fig:example_decision_trees} (a-2) and (b-2). Decision trees of depths 1 and 2 partition $\MX$ into two and four subspaces, with individuals whose $x$ belongs to each subspace assigned to one of the three options. Generally, a decision tree of depth $L$ partitions $\MX$ into $2^L$ subspaces.

\begin{center}
[Figure \ref{fig:example_decision_trees} about here]
\end{center}

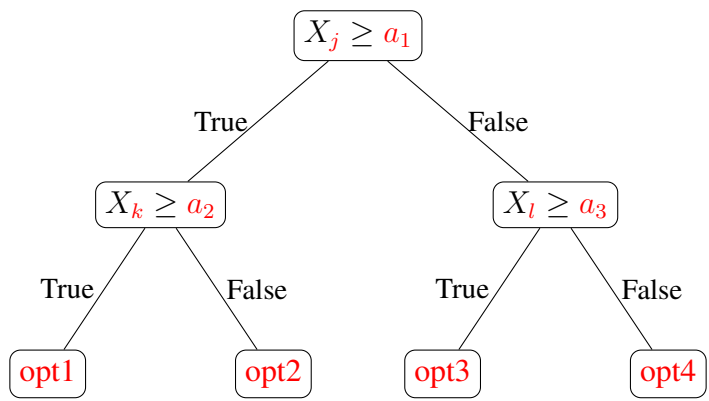
\begin{figure}[H]
    \centering
    \caption{Decision tree of depth 2}\label{fig:search_decision_tree}

    \begin{tikzpicture}
[   scale=1.5,
    level 1/.style = {sibling distance = 3.5cm},
    level 2/.style = {sibling distance = 2cm}
]
 
\node [draw,rounded corners] {\large $X_{\textcolor{red}{j}} \geq \textcolor{red}{a_{1}}$}
    child {node [draw,rounded corners] {\large $X_{\textcolor{red}{k}} \geq \textcolor{red}{a_{2}}$} 
    child {node [draw,rounded corners][draw,rounded corners]  {\large \textcolor{red}{\large opt1}} edge from parent node [left] { True}}
    child {node [draw,rounded corners] {\large \textcolor{red}{\large opt2}} edge from parent node [right] { False}}
    edge from parent node [left] { True}} 
    child {node [draw,rounded corners] {\large $X_{\textcolor{red}{l}} \geq \textcolor{red}{a_{3}}$}
    child {node [draw,rounded corners] {\large \textcolor{red}{\large opt3}} edge from parent node [left] { True}}
    child {node [draw,rounded corners] {\large \textcolor{red}{\large opt4}} edge from parent node [right] { False}}
    edge from parent node [right] { False}
    };
 
\end{tikzpicture}

\bigskip

    \begin{tablenotes}\footnotesize
\item[] Notes: $(j,k,l) \in \{1,\ldots,K\}^3$, $(a_1,a_2,a_3) \in \Real^3$, and $(\mbox{opt1},\ldots,\mbox{opt4}) \in \{T,NT,O\}^4$. Searching for the optimal decision tree of depth 2 is equivalent to finding the best combination of indices $(j,k,l) \in \{1,\ldots,K\}^3$ of $X$ and threshold values $(a_1,a_2,a_3) \in \Real^3$ in the top 2 layers, and options $(\mbox{opt1},\ldots,\mbox{opt4}) \in \{T,NT,O\}^4$ in the bottom layer.
\end{tablenotes} 
\end{figure}

A decision tree of depth $L$ consists of two components: (i) a set of inequalities allocated to the nodes in the top $L-1$ layers and (ii) a set of options allocated to the terminal nodes. Thus searching an optimal decision tree of depth $L$ corresponds to searching for an optimal combination of inequalities in the nodes in the top $L-1$ layers and an assignment option for each terminal node. 
For the example depth 2 decision tree in Figure \ref{fig:search_decision_tree}, searching for the optimal tree is equivalent to optimally choosing an $X$ for each node in the first and second layers (i.e., triplet of indices $(j,k,l) \in \{1,\ldots,K\}$ of the elements of $X$ where $K$ denotes the dimension of $X$), threshold values $(a_1,a_2,a_3)$ for these same nodes, and an assignment option $(\mbox{opt1},\ldots,\mbox{opt4}) \in \{T,NT,O\}^4$ for each of the bottom nodes. 

Learning an optimal decision tree of depth $L$ by the EWM method corresponds to finding a tree partition that maximizes the empirical welfare function $\MW(\hat{G})$ over a class $\MG$ of decision trees of depth $L$. The complexity of the policy class $\MG$ can be controlled by fixing the depth of possible decision trees (see, e.g., \cite{zhou_et_al_2018}).

\begin{center}
[Figure \ref{fig:search_decision_tree} about here]
\end{center}

\section{Field Experiment and Data}

To estimate an optimal arm assignment policy, we use data from a field experiment in Japan involving an energy savings program. Section \ref{sec:Field_Experiment} provides an overview of the field experiment. Section \ref{sec:Data} presents summary statistics and balance test results. Section \ref{sec:Heterogeneity} reports estimated average treatment effects and presents evidence of treatment effect heterogeneity. 

\subsection{Field Experiment}\label{sec:Field_Experiment}
Our field experiment was conducted in the summer of 2020 in collaboration with the Ministry of the Environment, Government of Japan. The population was households in the Kinki and Chubu regions of Japan. The experiment involved compulsory and opt-in schemes that provided monetary rewards for residential electricity saving. Ex-ante approval for the field experiment was obtained from the ethics committee of the Inter-Graduate School Program for Sustainable Development and Survivable Societies, Kyoto University.

In order to include a broad set of households, customers from 2 large regional power companies and 6 start-up\footnote{Entry into the Japanese retail electricity sector was fully liberalised in 2016. These 6 companies are post-2016 entrants.} power companies were invited to participate by letter or email. 
 The homes of customers of these power companies are equipped with advanced meters, known as “smart meters,” which record their electricity usage in 30-minute intervals. A total of 4,446 customers pre-registered for the experiment. Among them, those who canceled their electricity contracts in the middle of the experiment, those whose electricity usage data were not accurately recorded in their home energy management systems, and office customers who used a large amount of electricity, unlike ordinary households, were excluded. This left us with 3,870 households as participants of the entire experiment.

We randomly assigned each of the 3,870 households to one of three groups: No-treatment ($NT$), Compulsory treatment ($T$), and Opt-in treatment ($O$).\footnote{The random assignment process was designed so that $NT$: $T$: $O$= 2: 2: 1. A relatively large number of households were assigned to the $NT$ and $T$ groups as the data for these groups will also be used in future studies in this project.} All participating customers agreed to provide their electricity usage data and received a participation reward of 2,000 JPY (approximately 19 USD, given 1 USD = 105 JPY in the summer of 2020)

\vskip\baselineskip

{\bf No-treatment ($NT$)}: The 1,577 customers in this group did not receive the reward treatment.

{\bf Compulsory treatment ($T$)}: The 1,486 customers in this group received monetary rewards for their electricity savings.

{\bf Opt-in treatment ($O$)}: The 807 customers in this group were given the option to complete an application process in advance and receive the treatment. The opt-in rate, was 37.17\% (300 out of 807 customers).

\vskip\baselineskip

This experiment aimed to reduce residential electricity consumption between 1:00 and 5:00 pm during the week of August 24 to 30.  The 1:00 to 5:00 pm period is when electricity demand peaks in the summer in Japan, and is known as "peak hours". We set the baseline for each household to be their peak-hour electricity consumption in July. Households in the Compulsory treatment group, and those in the Opt-in treatment group who completed the application process, were provided with a monetary reward of 100 JPY per 1 kWh reduction in their peak-hour electricity consumption relative to their baseline. The reward payment for a household was calculated as follows. Let $\bar{Y}_i^{base}$ denote the baseline, which we set to be average daily peak-hour electricity consumption (kWh) from July 1 to 31. Let $\bar{Y}_i^{treat}$ be 
the average daily peak-hour electricity consumption (kWh) from August 24 to 30. We define the average daily peak-hour electricity saving, $\Delta \bar{Y}_i$, to be the difference between $\bar{Y}_i^{treat}$ and $\bar{Y}_i^{base}$ truncated to be positive, so that $\Delta \bar{Y}_i = - {\rm min} \{0, \bar{Y}_i^{treat} - \bar{Y}_i^{base}\}$. We multiply the average daily of peak-hour electricity saving, $\Delta Y_i$, by seven days, and define this to be the total energy savings for the entire week: $\Delta Y_i^{total}$. The total reward for household $i$, $Q_i$, is calculated as: 
\begin{eqnarray}
    Q_i &=& \Delta Y_i^{total} \times 100 \nonumber \\
        &=& -{\rm min} \{0, \bar{Y}_i^{treat} - \bar{Y}_i^{base}\} \times 100 \nonumber
\end{eqnarray}

Information regarding the treatment week, peak hours, and reward calculation procedure was delivered to the Compulsory-treatment and Opt-in treatment groups by letter or email on July 31. Customers assigned to the Opt-in treatment group who wished to receive the reward were required to complete an application process in advance. Customers in this group could complete this process by returning a prepaid postcard\footnote{Specifically, what is known in Japan as a "round-trip postcard."} or online during the two-week period from July 31 to August 11. 300 out of 807 customers  (37.17\%) in this group completed the process and expressed their willingness to receive the treatment. This opt-in rate is slightly higher than those for Critical Peak Pricing (CPP) in previous studies, 20\% \citep{Potter_et_al_2014, Fowlie_et_al_2021} and 16-31\% \citep{Ito_2019}. Unlike the CPP treatment, where the price of electricity can increase, the reward amount in this experiment will never be negative, even a customer's peak-hour electricity consumption is above their baseline. This feature may have contributed to the high opt-in rate.

The reward calculation procedure was described as follows: “we will pay you a reward of 100 JPY per 1 kWh of electricity saved when you pay your electricity bill,” and “we calculate your electricity savings rate using your total electricity consumption during these specific hours and your average electricity consumption in the previous month.” In general, experiments featuring reward treatments need to address concerns regarding baseline manipulation by customers \citep{Wang_2018}. Customers who are aware how the reward is calculated have an incentive to increase their electricity consumption during the baseline period in order to obtain a greater reward. This baseline manipulation is often observed in studies using reward treatments and could be critical to the estimation of treatment effects \citep{Wolak_2007}. However, such manipulation was difficult in this experiment as the baseline period was a month long, and participants were only made aware of the calculation process on the final day of this month. Table \ref{tab:summary} in the following subsection shows no difference in baseline period electricity consumption between the three groups, consistent with a lack of manipulation.

\subsection{Data and Summary Statistics}\label{sec:Data}

We used high frequency data of electricity consumption and response data from a survey conducted upon obtaining customers' agreement to participate in the experiment. Columns 1, 2, 3 of Table \ref{tab:summary} present summary statistics for pre-experiment consumption data and demographic variables by group. The variables are pre-experiment peak hour-usage, pre-experiment pre-peak hour-usage, pre-experiment post-peak hour-usage, number of people at home on weekdays, interest in energy conservation, and household income.\footnote{The variable peak hour-usage is the average electricity consumption per 30-minutes during peak hours (from 1:00 to 5:00 pm), the variable pre-peak hour-usage is the average electricity consumption per 30-minutes during pre-peak hours (from 10:00 am to 1:00 pm), and the variable post-peak hour-usage is the average electricity consumption per 30-minutes during post-peak hours (from 5:00 to 8:00 pm).} We observe no significant difference in the first five variables between groups. This indicates that random assignment has statistically balanced these five observables. For household income, we observe no significant difference between the $NT$ and $O$ groups, or the $T$ and $O$ groups, but we do observe a significant difference between the $NT$ and $T$ groups. However, we confirmed that the estimated treatment effects did not substantially vary when controlling for demographic variables including household income.

\begin{center}
[Table \ref{tab:summary} about here]
\end{center}

We investigated the external validity of our sample by randomly sampling 2,070 customers who did not participate in this experiment from the same population and comparing this random sample with our sample. We found that our sample has larger pre-experiment electricity usage, a larger number of people at home on weekdays, higher interest in energy conservation, and higher household income. The details are described in Appendix \ref{sec:externalvalidity}.

\subsection{Experimental Results and Heterogeneity}\label{sec:Heterogeneity}
Average treatment effects are estimated using peak-hour electricity consumption data recorded for 30-minute intervals and the following equation:
\begin{eqnarray}
    \log{Y_{it}} = \sum_{d=\{T, O\}}  \tau^{ITT}_d Z^d_{it} + \lambda_i +\theta_t + \epsilon_{it}
\end{eqnarray}
where $log{Y_{it}}$ the natural log of electricity usage for household $i$ in a 30-minute interval $t$. We included household fixed effects $\lambda_i$, and time fixed effects $\theta_t$ for each 30-minute interval to control for time-specific shocks such as weather. $\epsilon_{it}$ is an unobservable error term assumed to follow a normal distribution with mean zero. $Z^d_{it}$ equals one if household $i$ is in the Compulsory treatment or the Opt-in treatment in $t$. $\tau^{ITT}_d$ represents the treatment effect for each group. In particular, $\tau^{ITT}_O$ represents the Intention-to-Treat (ITT) estimate, since households in the Opt-in group are required to apply to receive the reward treatment.

\begin{center}
[Table \ref{tab:ITT} about here]
\end{center}

Column 1 of Table \ref{tab:ITT} shows that the Compulsory treatment led, on average, to a reduction in peak-hour electricity consumption for the summer treatment week of 0.097 log points and the Opt-in treatment caused a reduction in consumption of 0.052 log points. The former effect is slightly larger than the latter, and the difference is statistically significant at the 10\% level.

The remaining columns of Table \ref{tab:ITT} investigate heterogeneity in the treatment effects. Each pair of columns splits the sample into two groups: those with a below median value of a particular variable, and those with an above median value. The variables used in splitting are the difference between peak-hour and pre-peak-hour average electricity consumptions during the baseline period, the difference between peak-hour and post-peak-hour average electricity consumptions during the baseline period, the number of people at home on weekdays, interest in energy conservation, and household income.

Focusing on the difference between peak-hour and pre-peak-hour average electricity consumption during the baseline period, the ITT estimate for the lower group is -0.108 for Compulsory treatment and -0.022 for Opt-in treatment. In contrast, the estimates for the upper group are -0.079 and -0.073. The $p$-value for the test of difference between the two ITT estimates is 0.013 for the lower group and 0.880 for the upper group. That is, the Compulsory treatment promotes electricity saving more effectively than the Opt-in treatment for households with a smaller difference between pre-experiment peak-hour and pre-peak-hour average electricity consumption. However, there is no such statistically significant difference for the other sub-group, which implies there is heterogeneity in the relative effectiveness of the two treatments between the two sub-groups.

We also observed statistical differences between the two ITT estimates for some other sub-groups, including the households with a smaller difference between pre-experiment peak-hour and post-peak-hour average electricity consumptions, those with a smaller number of people at home on weekdays, those with a lower interest in energy conservation, and those with a higher household income. This implies that the relative effectiveness of the Compulsory treatment and the Opt-in treatment can vary substantially depending on households' pre-experiment electricity usage and demographics, which in turn suggests that the optimal treatment assignment may differ across households. 

\section{Optimal Assignment Policy and Welfare Gain: Empirical Results \label{sec:optimal policy design}}

This section estimates optimal assignment policies using the data from our RCT, and compares the optimal paternalistic assignment with autonomous choice and a mixture of the two approaches. This analysis leads us to two findings: (i) paternalistic assignment and aunotonomous choice have similar performance, (ii) mixing the two improves welfare significantly. We then present empirical evidence on the mechanism driving this gain in welfare. 

\subsection{Construction of Social Welfare Criterion}\label{sec:construction-of-social-welfare-criterion}

We start by setting up a social welfare criterion. Our primary interest is in learning how the total surplus changes when moving from the status quo (no energy saving program) to an individualized energy saving program. We hence set the total surplus as the social welfare criterion to be maximized. 

For each $j\in\{T,NT,O\}$, let $Y(j)$ be a household's potential electricity consumption (kWh) for the experiment period, and $Z(j)\in\{T,NT\}$, $j\in\{T,NT,O\}$ be a potential choice variable indicating the treatment the household would receive if assigned to arm $j$. Since noncompliance is not possible for the compulsory arms, $Z(j)=j$ holds for $j=T$ and $NT$. On the other hand, if a household were assigned to the opt-in treatment arm, $Z(O)$ indicates their choice of treatment.

We define four parameters: $p$, $c$, $a$, and $\delta$. $p$ is the unit price of electricity. We set to $p=25$ JPY/kWh, approximately the regulated price of electricity in Japan, which is independent of the time of a day.\footnote{In Japan, until April 1 2016 household electricity was supplied by local power companies and retail prices were regulated. Since then, entry into the retail electricity industry has been fully liberalized, allowing all households to freely choose their price menu. However, as a transitional measure, the regulated price for households is being maintained for the time being, and is set at approximately 25 JPY/kWh regardless of the time of day.} $c$ is the marginal cost of production for electricity. We specify $c=125$ JPY/kWh, so that the difference between $p$ and $c$ is equal to the per kWh rebate. The wholesale price of electricity sometimes soars during peak hours such as summer afternoons or winter evenings, reflecting supply constraints. In the past, the electricity price has occasionally exceeded 100 JPY/kWh in summer afternoons.\footnote{The wholesale electricity market, where the power generation sector and the retail sector trade electricity, is operated by the Japan Electric Power Exchange (JEPX). Most trading takes place in the ``day-ahead market'' where both sectors trade electricity on the day before the actual demand period. Trading results are disclosed, and we confirm that the price exceeded 100 JPY/kWh on July, 25, 2018. Moreover, the price has exceeded even 125 JPY/kWh. For example, it the price reached 250 JPY/kWh on January, 15, 2021.}

Parameter $a$ represents the administrative cost of implementing our energy saving program. This cost is comprised of several items, including the installation cost of the Home Energy Management System (HEMS) required to participate. In 2016, the Japanese government estimated the cost of implementing a demand reduction program, including the installation cost of HEMS, to be 291.1 JPY/kWh per household per season \citep{Ida2017}.\footnote{We do not include the installation cost for a smart meter in the administrative cost. Since the Great East Japan Earthquake of March 11, 2011 and the accident at the Fukushima Daiichi Nuclear Power Plant, the Japanese government has stipulated that smart meters be installed in all homes by the end of the decade, so this cost is ``sunk'' in that it will be paid regardless of whether a demand reduction program is implemented or not.} We use this as the value of the administrative cost.

Parameter $\delta$ represents the long-term benefits of a unit reduction in energy consumption. Here, we consider the effect of a unit reduction on the capacity market, where future supply capacity is traded between the power generation and retail sectors. In Japan, the capacity market was established in 2020, with the first auction held at that time. In that auction, the Japanese government provided a reference price 9,425 JPY/kW to bidders. We use this value divided by the total event hours (28 hours) as the value of $\delta$.

With these variables and parameters, we specify the social welfare of assignment policy $G$ to be
\begin{align}
\begin{split}
    \mathcal{W}(G):={} & E\left[\left\{ \left(\delta+\frac{p-c}{2}\right)(Y(T)-Y(NT))-a\right\} \cdot1\{X\in G_{T}\}\right.\\
    & \qquad+\left.\left\{ \left(\delta+\frac{p-c}{2}\right)(Y(O)-Y(NT))-a\cdot1\{Z(O)=T\}\right\} \cdot1\{X\in G_{O}\}\right]
\end{split}
\label{eq:definition-of-surplus}\\
\begin{split}
    ={} & E\left[\sum_{j\in\{T,NT,O\}}\left\{ \left(\delta+\frac{p-c}{2}\right)Y(j)-a\cdot1\{Z(j)=T\}\right\} \cdot1\{X\in G_{j}\}\right]\\
    & \qquad-E\left[\left(\delta+\frac{p-c}{2}\right)Y(NT)\right].
\end{split}
\label{eq:equivalent-representation-of-surplus}
\end{align}
Equation~\eqref{eq:definition-of-surplus} gives our definition of the change in total surplus. For each $j\in\{T,NT,O\}$, the term $\frac{p-c}{2}(Y(j)-Y(NT))$ represents the household's contribution to the short-term welfare when it is assigned to arm $j$, while the term $\delta(Y(j)-Y(NT))$ measures the household's contribution to long-term welfare. In addition, $a\cdot1\{Z(j)=T\}$ represents the administrative cost of implementing the energy saving program. We here assume that this cost is realised when the household actually takes up the program. In total, the term $(\delta+\frac{p-c}{2})(Y(j)-Y(NT))-a \cdot 1\{Z(j)=T\}$ measures the individual welfare contribution net of the treatment cost. Equation~\eqref{eq:equivalent-representation-of-surplus} reveals that setting $W(j)=(\delta+\frac{p-c}{2})Y(j)-a\cdot1\{Z(j)=T\}$ allows us to apply the empirical welfare maximization framework described in Section~2. Note that $\mathcal{W}(G)=0$ holds for the uniform compulsory no-treatment policy; i.e., $G=(G_{T},G_{NT},G_{O})=(\emptyset,\mathcal{X},\emptyset)$. Hence, the social welfare of policy $G$ can be interpreted as the welfare gain relative to a policy of uniform no-treatment. 

We implement the EWM method using the social welfare function defined above. As stated in Section~3, our sample consists of 3,870 households. Let $Y_{i}$ denote household $i$'s observed electricity consumption level (kWh) during the treatment period (i.e., from 1:00 to 5:00 pm on August 24 to 30), and $Z_{i}\in\{T,NT\}$ denote household $i$'s observed decision about over participation in the energy saving program. As with the potential choice variables introduced earlier, we have $Z_{i}\equiv T$ or $Z_{i}\equiv NT$ when the household is assigned to $T$ or $NT$. On the other hand, when the household is assigned to $O$, the value of $Z_{i}$ changes depending on the household's decision. We estimate the optimal policy $\widehat{G}$ by maximizing the following objective function over a class of policies $\mathcal{G}$:
\[
\frac{1}{n}\sum_{i=1}^{n}\sum_{j\in\{T,NT,O\}}\frac{W_{i}\cdot1\{D_{i}=j\}}{P(D_{i}=j\mid X_{i})}\cdot1\{X_{i}\in G_{j}\},
\]
where
\[
W_{i}:=\text{\ensuremath{\left(\delta+\frac{p-c}{2}\right)}}Y_{i}-a\cdot1\{Z_{i}=T\}.
\]
That is, we obtain an optimal policy $\widehat{G}$ by maximizing the sample analog of the first term in \eqref{eq:equivalent-representation-of-surplus}. Since the second term in \eqref{eq:equivalent-representation-of-surplus} does not depend on the policy $G$, this is equivalent to maximizing the sample analog of $\mathcal{W}(G)$.

During estimation, we let $Y_{i}$ be the difference between the observed electricity consumption level during the treatment period and the average consumption level during the baseline period (i.e., from 1:00 to 5:00 pm of July 1 to 31). By defining $Y_i$ as the consumption level relative to a baseline, we can control for an additive household specific unobservable without changing the optimal assignment policies for the population. Eliminating this household specific unobservable results in a more accurate estimate of the empirical welfare criterion. Since the baseline consumption level is observed before random assignment, this operation does not affect the expected welfare value. We demean the observed outcome $Y_{i}$ as suggested in \citet{KT18}. In addition, we replace the propensity score $P(D_{i}=j\mid X_{i})$ with the sample fraction, i.e., $\sum_{i=1}^{n}1\{D_{i}=j\}/n$.

We specify the policy class to be the class of decision trees of a depth of either 3 or 6. We select five variables to be used in constructing the decision trees. These are Peak - Pre-peak, Peak - Post-peak, household income, the number of residents in the same housing unit, and a measure of the households willingness to participate in the energy saving program. The first four variables are selected based on their ability to predict electricity consumption and the conditional average treatment effects. Specifically, we select these four variables by running two off-the-shelf machine learning algorithms, lasso and random forest, with all the available covariates and assessing the importance of each variable.
When using lasso to assess importance, we regress $Y_{i}$ on all the available covariates with a $l_{1}$-penalization term. We order variables in terms of importance by increasing the penalization parameter step-wise and checking which variables remain selected for large penalization parameter values. When using random forest, we estimate the conditional average treatment effects using the causal forest algorithm of \citep{Wager2018} with all available covariates included. We use the frequency with which a variable is used to split nodes as a measure of its importance. The four selected variables are those that appeared on the lists of important variables produced by both methods. The fifth variable, the willingness to participate in the energy saving program, was included because it is expected to be useful for predicting the opt-in decision. We believe this is reasonable since the opt-in decision plays an important role in our social welfare function, as is clear from equation~\eqref{eq:definition-of-surplus}.



\subsection{Empirical Results on Paternalistic Assignment Policy}\label{sec:empirical-results-on-paternalistic-assignment}

We first focus on comparing paternalistic assignment(assigning either $T$ or $NT$ according to household characteristics) with autonomous choice (uniform assignment of $O$). We show estimates of the welfare gains relative to uniform assignment of $NT$ to compare their performance in terms of social welfare.

To find the optimal paternalistic assignment policy, we maximize the estimates of the social welfare over the class of decision tree policies, with the arms restricted to $T$ and $NT$. We set the decision tree depth to 3, and exactly maximize the empirical welfare criterion by applying the exhaustive search algorithm of \cite{zhou_et_al_2018}. We constrain the depth to 3 since, given our sample size, obtaining an exact globally optimal tree of depth greater than 3 in reasonable time is difficult. We also obtain a (bias-corrected) estimate of the welfare gain of the optimal paternalistic assignment policy and compare it with the welfare gain of a policy of uniform autonomous choice.  

\begin{figure}[H]
\centering
\caption{Optimal paternalistic assignment policy}
\includegraphics{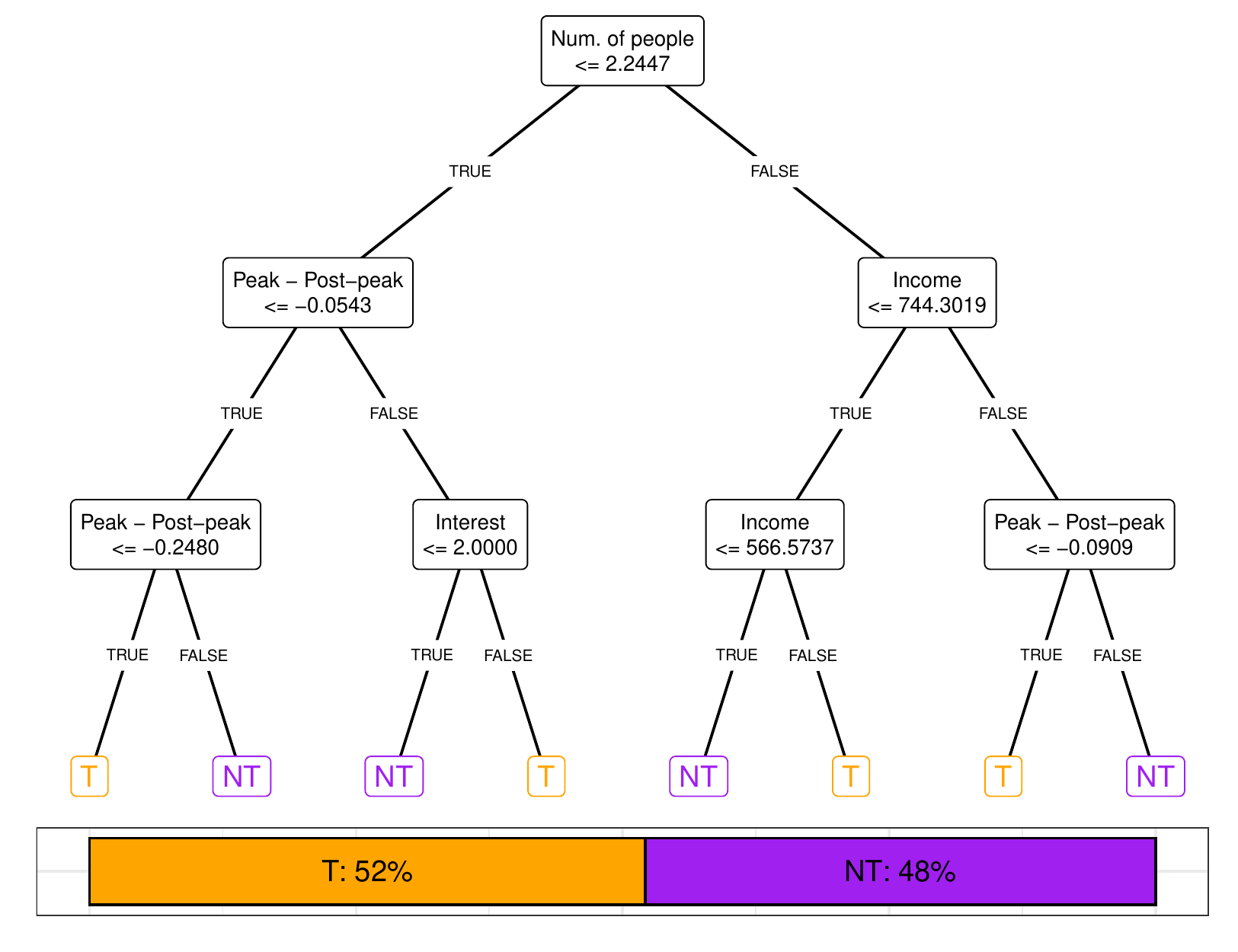}
\label{fig:tree-two-option}
\begin{tablenotes}[flushleft]
    \footnotesize\item[] Notes: This figure shows the optimal paternalistic policy estimated in Section~\ref{sec:empirical-results-on-paternalistic-assignment}. The upper panel illustrates the structure of the policy. By following the yes-no questions from the top of the tree down, each household is assigned to a specific arm. The lower panel shows the estimated share of households assigned to each arm under this policy.
\end{tablenotes}
\end{figure}

The upper panel of figure~\ref{fig:tree-two-option} presents the decision tree corresponding to the estimated assignment policy $\hat{G}^{pat} \equiv (\hat{G}_T^{pat}, \hat{G}_{NT}^{pat}, \emptyset)$. All variables except Peak - Pre-peak appear at least once in the tree. In particular, Peak - Post-peak is often used. The lower panel of figure~\ref{fig:tree-two-option} shows the share of households assigned to each arm at $\hat{G}^{pat}$. These are estimated as the fractions of households satisfying $X \in \hat{G}_T^{pat}$ and $X \in \hat{G}_{NT}^{pat}$ in the experimental data. 
$\hat{G}^{pat}$ assigns half of the households to compulsory enrollment in the energy saving program and excludes the other half, highlighting heterogeneity in the sign of welfare contributions among the households. As show below, this leads paternalistic assignment to dominate uniform assignment to treatment and no-treatment. 

\begin{center}
[Figure \ref{fig:tree-two-option} about here]
\end{center}

To assess the welfare performance of the estimated policy, we report point estimates and 95\% confidence intervals for the welfare gain relative to uniform assignment to $NT$. If an optimal assignment policy is estimated by EWM, the optimized empirical welfare value  will be an upwardly biased estimate of the true welfare attained by the estimated policy. This is known as the winner's bias (see, e.g., \citep{Andrews_etal_2019}), and is caused by using the same data twice: once to learn the policy and once to infer the policy's welfare.\footnote{The estimation and inference procedures proposed by \cite{Andrews_etal_2019} cannot be directly applied to decision tree based policies because the number of candidate policies is infinite.} To control for the winner's bias in our point estimates and confidence intervals, we create artificial test data by fitting a causal forest \cite{Wager2018} to run regressions of the outcome onto all the covariates, and generating data with permuted regression residuals. See the Appendix for the details. Using this artificial test data, we compute point estimates and confidence intervals for the welfare gains.

Table~\ref{tab:purely-paternalistic-approach} reports the welfare performances of the uniform benchmark policies (100\% $T$, 100\% $NT$, and 100\% $O$) and the estimated optimal paternalistic assignment policy $\hat{G}^{pat}$. The estimated welfare gain of uniform treatment (100\% $T$ assignment) is 63.1 JPY per household, which is not statistically significant in spite of the significant decrease of electricity consumption shown in section~3. 
This is due to the administrative cost of treatment; for households who respond little to the program, the treatment cost exceeds the benefit, and the net welfare contribution is negative. 
This is consistent with the large welfare improvement under the paternalistic assignment policy $\hat{G}^{pat}$. The estimated welfare of the uniform autonomous choice policy (100\% $O$) is 140.9 JPY per household, and this is also not statistically significant. Furthermore, the difference between 100\% $T$ and 100\% $O$ is not statistically significant ($p$-value is 0.39), implying that there is no obvious winner among the three uniform policies.

\begin{center}
[Table~\ref{tab:purely-paternalistic-approach} about here]
\end{center}

The fourth row shows the welfare performance of the estimated optimal paternalistic assignment policy. The estimated welfare is 228.5 JPY, which is significantly different from zero. That is, the paternalistic assignment policy improves welfare compared to the status quo policy (100\% $NT$). The remaining rows show the comparison of the paternalistic assignment policy and the other benchmark policies. The estimated welfare gain of the paternalistic policy compared to 100\% $T$ is 165.5 JPY per household, and this difference is statistically significant. 
This finding implies that the paternalistic policy exploits the heterogeneity in the sign of the net welfare contribution over the observable household characteristics. The final comparison is between the optimal paternalistic assignment policy and autonomous choice. The estimated welfare gain is 87.6 JPY per household, and this difference is not statistically significant. That is, we find no evidence that one approach strictly dominates the other. These results lead to our next empirical question: whether a mixture of paternalistic and autonomous approaches can outperform both alone.

\subsection{The Optimal Mix of Paternalism and Autonomy}\label{sec:optimal-mix-of-paternalism-and-autonomy}

Having observed the similar welfare performance of paternalistic assignment $\hat{G}^{pat}$ and autonomous choice, we now consider whether or not performance can be improved by a targeted mixing of paternalistic assignment and autonomous choice. 
In our framework, a policy assigning arms $NT$, $T$, or $O$ in response to observable household characteristics represents a mixture of paternalistic assignment and autonomous choice. Similar to the estimation of the optimal paternalistic assignment policy, we search for an optimal policy in a class of decision trees, now with one of these three possible arms assigned to each terminal node. To allow for a more flexible policy in line with the increase in available options, we set the tree depth to 6. This comes with a computational cost that prevents us from computing the exact global maximizer of empirical welfare. The exhaustive search algorithm for decision tree of depth 3, as used to estimate the optimal paternalistic assignment policy, completes within a reasonable time frame, but increasing the depth beyond 3 renders exhaustive search infeasible. 

We therefore consider a heuristic search procedure to approximately solve the empirical welfare maximization problem. We divide the search for an optimal tree partition into two steps. In the first step, we find an optimal tree of depth 3. In the second, we search within each leaf obtained in the first step for an optimal subtree of depth 3. The resulting partition is a decision tree of depth 6, but is not guaranteed to maximize the empirical welfare over the class of decision trees with depth 6. See the Appendix for further details. 
We denote a policy estimated by this two-step procedure by $\hat{G}^{mix}$. Since $\hat{G}^{mix}$ is not necessarily a global maximizer of empirical welfare, the welfare estimate at $\hat{G}^{mix}$ can be lower than the welfare estimate at a globally optimal policy.
Therefore, the welfare gain estimates reported below can be viewed as conservative estimates of the welfare gain attained by an optimal targeting policy that mixes paternalistic assignment and autonomous choice.

\begin{figure}[H]
\caption{Optimal mixture of the paternalistic assignment and autonomous choice: from depth 1 to 3}
\centering
\includegraphics{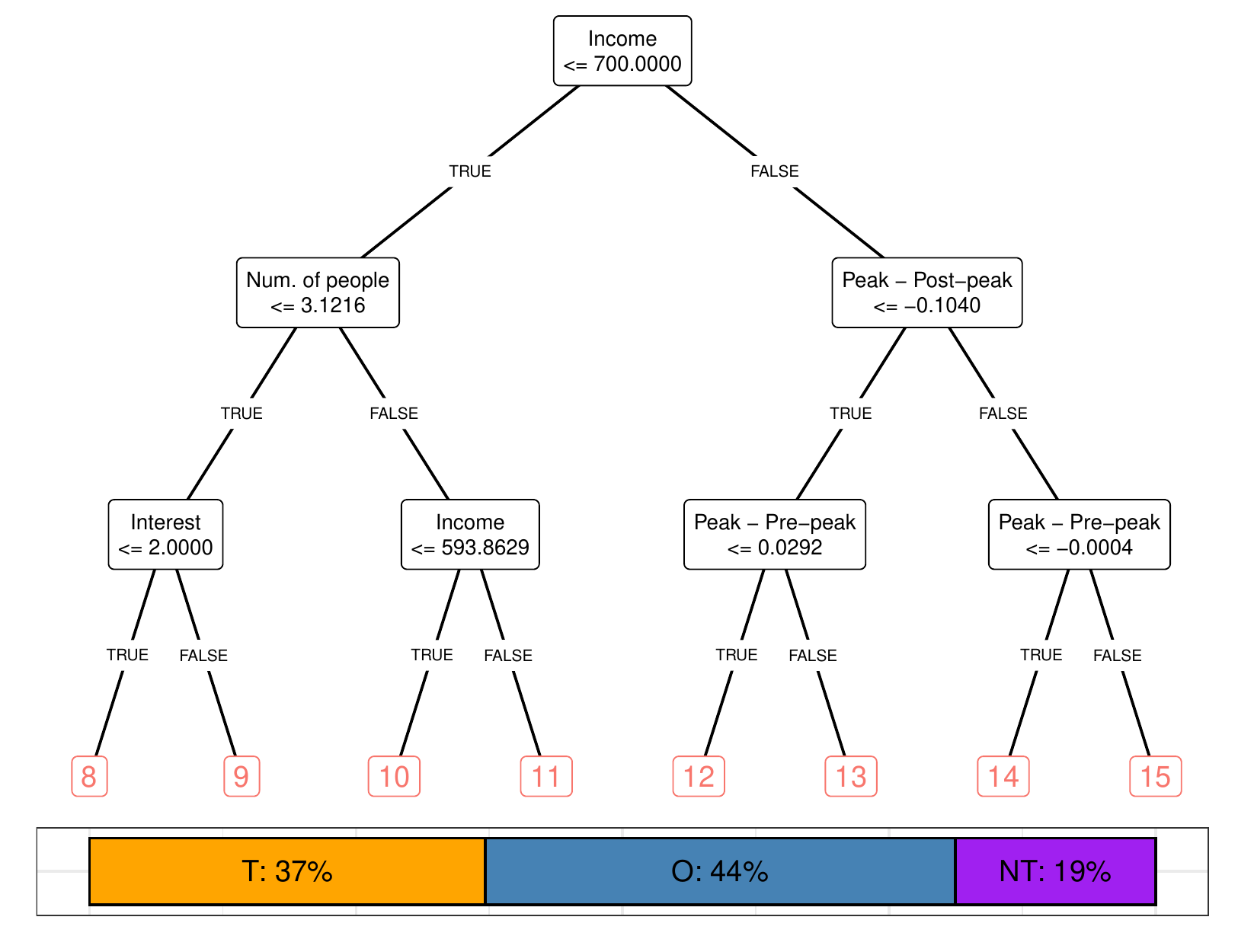}
\label{fig:tree-three-option-1-3}
\begin{tablenotes}[flushleft]
    \footnotesize \item[] Notes: This figure shows the optimal mixture of the paternalistic assignment and autonomous choice, which is estimated in Section~\ref{sec:optimal-mix-of-paternalism-and-autonomy}. The upper panel illustrates the upper half of the structure of the policy. By following the yes-no questions from the top of the tree down, each household is assigned a number. If the household is assigned a number less than or equal to 11, the household proceeds to figure \ref{fig:tree-three-option-4-6-left}; otherwise, the household proceeds to figure \ref{fig:tree-three-option-4-6-right}. The lower panel shows the estimated share of households assigned to each arm under this policy.
\end{tablenotes}
\end{figure}

\begin{figure}[H]
\caption{Our best mixture of paternalistic and autonomous approaches: from depth 4 to 6, right side}
\centering
\includegraphics[scale=0.5]{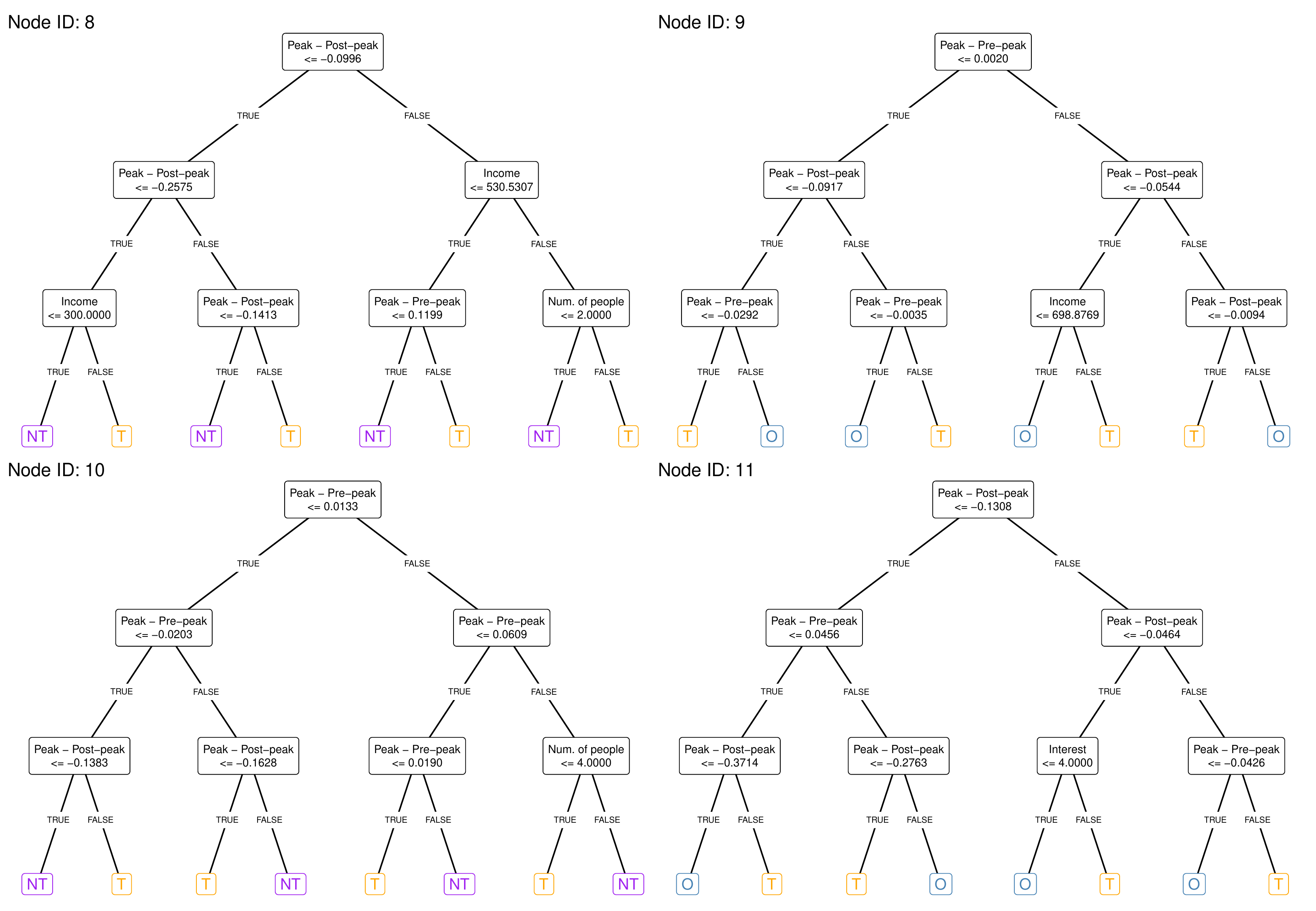}
\label{fig:tree-three-option-4-6-left}
\begin{tablenotes}
    \footnotesize\item[]Notes: This figure shows the optimal mixture of the paternalistic assignment and autonomous choice, which is estimated in Section~\ref{sec:optimal-mix-of-paternalism-and-autonomy}. Specifically, it illustrates the lower left section of the decision tree. Households assigned numbers between 8 and 11 in figure \ref{fig:tree-three-option-1-3} are assigned to arms by working down through the yes-no questions in the tree corresponding to their number. 
\end{tablenotes}
\end{figure}

\begin{figure}
\caption{Our best mixture of paternalistic and autonomous approaches: from depth 4 to 6, left side}
\centering
\includegraphics[scale=0.5]{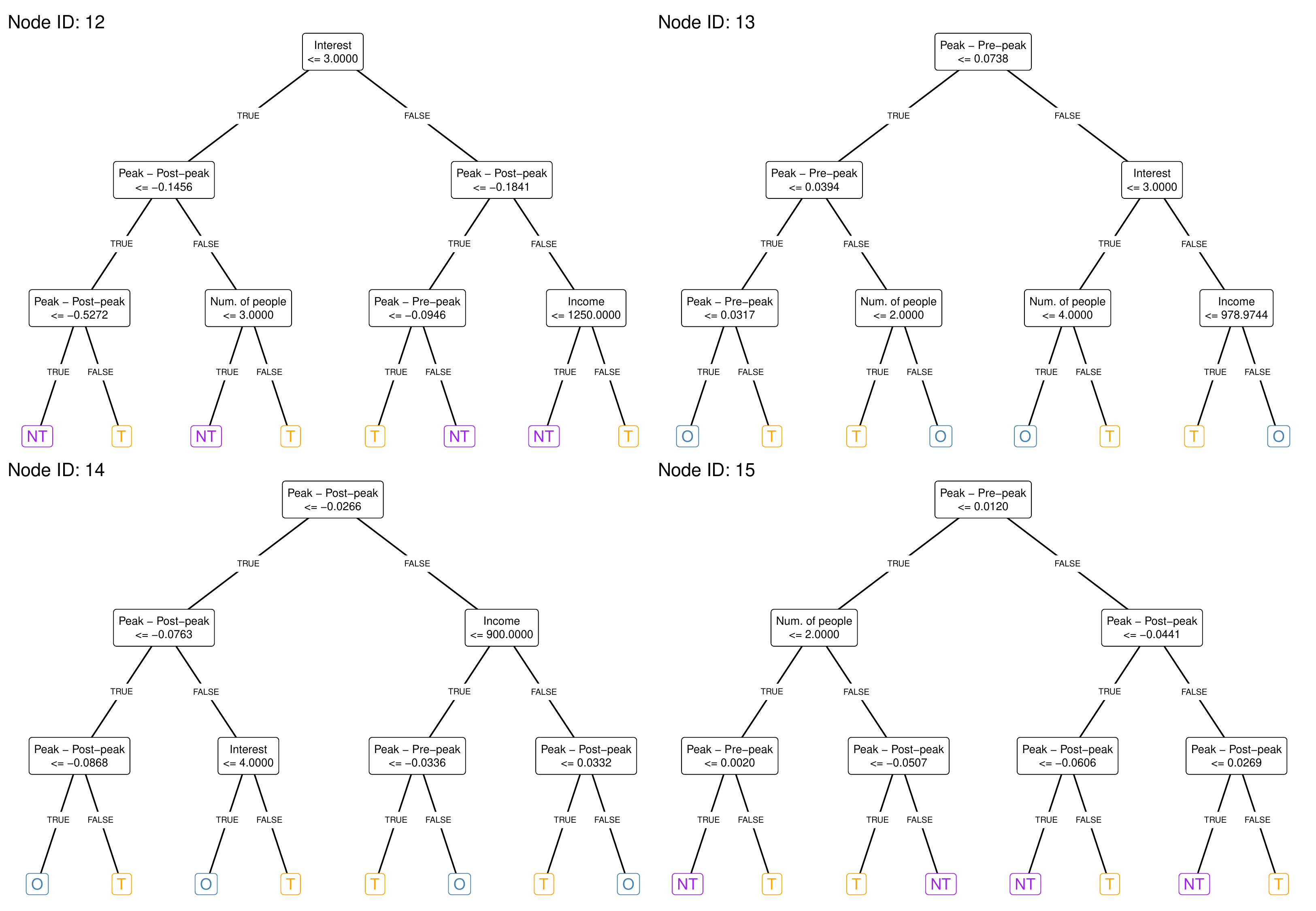}
\label{fig:tree-three-option-4-6-right}
\begin{tablenotes}
    \footnotesize\item[]Notes: This figure shows the optimal mixture of the paternalistic assignment and autonomous choice, which is estimated in Section~\ref{sec:optimal-mix-of-paternalism-and-autonomy}. Specifically, it illustrates the lower right section of the decision tree. Households assigned numbers between 12 and 15 in figure \ref{fig:tree-three-option-1-3} are assigned to arms by working down through the yes-no questions in the tree corresponding to their number. 
\end{tablenotes}
\end{figure}

Figures~\ref{fig:tree-three-option-1-3},\ref{fig:tree-three-option-4-6-left}, and \ref{fig:tree-three-option-4-6-right} show the estimated assignment policy. The policy assigns households to arms as follows. First, the binary conditions of figure \ref{fig:tree-three-option-1-3} assign each household to one node between 8 and 15. Say a household is assigned to node 8, then this household will be assigned to an arm according to the additional binary conditions for node 8 presented in figure \ref{fig:tree-three-option-4-6-left}. Figure \ref{fig:tree-three-option-1-3} also shows the share of households assigned to each arm. Under the mixed policy, 44\% of the households are assigned to arm $O$, 37\% to arm $T$, and 19\% to arm $NT$. In comparison to the paternalistic assignment policy estimated previously, $\hat{G}^{mix}$ assigns lower shares of households to $T$ and $NT$, with almost half of them assigned to arm $O$.

\begin{center}
[Figure \ref{fig:tree-three-option-1-3}, \ref{fig:tree-three-option-4-6-left}, and \ref{fig:tree-three-option-4-6-right} about here]
\end{center}

\begin{figure}[htpb]
\centering
\caption{Two dimensional summary of the policies}
\includegraphics[scale=0.65]{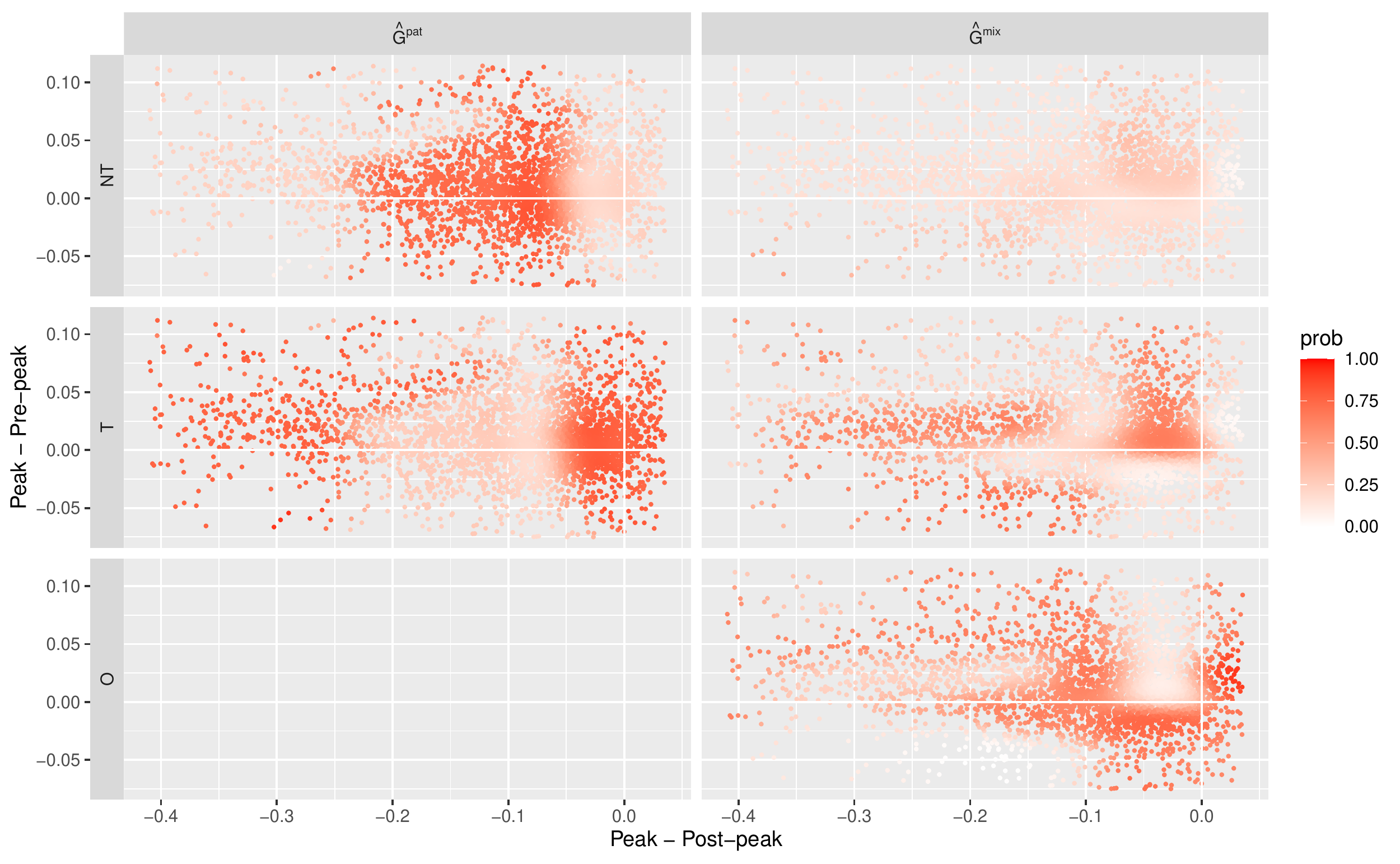}
\label{fig:two-dimensional-summary}
\begin{tablenotes}
    \footnotesize\item[]Notes: This figure shows a two dimensional summary of the policies estimated in Section~\ref{sec:empirical-results-on-paternalistic-assignment} and \ref{sec:optimal-mix-of-paternalism-and-autonomy}. Each row corresponds to a different arm, and each column to a different policy ($\hat{G}^{pat}$ or $\hat{G}^{mix}$). In each panel, a point represents a household observed in our experimental data, and the color of the point shows the probability of being assigned to a specific arm.
\end{tablenotes}
\end{figure}

The decision tree is sufficiently complex that its overall structure is difficult to grasp. To summarize the assignment policy, we regress the assigned arm on two key covariates, Peak - Pre-peak and Peak - Post-peak, and show how assignment probabilities vary with respect to them. Figure~\ref{fig:two-dimensional-summary} summarizes the probability estimates. Panels on the left show assignment probabilities for the optimal paternalistic policy, panels on the right show assignment probabilities for the mixed policy. Each point represents a household observed from our sample, and the color of the point shows the probability of being assigned to a specific arm. Under the paternalistic policy, assignment is mainly determined by Peak - Post-peak. On the other hand, under the mixed policy, Peak - Pre-peak also plays a role in determining assignment.

\begin{center}
[Figure \ref{fig:two-dimensional-summary} about here]
\end{center}

Table~\ref{tab:mixture-approach} presents the estimated welfare performance of $\hat{G}^{mix}$ and the benchmark policies. The welfare gain of $\hat{G}^{mix}$, shown in the fifth row, is 437.9 JPY and is significantly different from zero. Rows six through eight present comparisons of the welfare gain from the mixed policy with other policies. 
There are two important findings. First, $\hat{G}^{mix}$ attains higher welfare than autonomous choice. The estimated welfare gain relative to autonomous choice is 297.0 JPY, and the 95\% confidence interval does not contain zero. Second, $\hat{G}^{opt}$ outperforms the paternalistic assignment policy $\hat{G}^{pat}$. The difference of 209.4 JPY, which is significantly different from zero. We can thus conclude that proper targeting of paternalistic assignment and autonomous choice based on observable household characteristics can significantly improve social welfare in comparison to either policy alone. 

\begin{center}
[Table \ref{tab:mixture-approach} about here]
\end{center}

\subsection{What Delivers Welfare Gain?}\label{sec:what-delivers-welfare-gain}

In this section, we investigate the mechanism driving the welfare gain from the mixed policy $\hat{G}^{mix}$. In particular, we focus on the average welfare effect for subgroups of households who are assigned to arms that they would not autonomously choose by $\hat{G}^{mix}$.   

For a given assignment policy $G=(G_{T},G_{NT},G_{O})$, the RCT data can be used to identify the average welfare gain relative to $W(NT)$ conditional on this assignment policy $G$, i.e., $E[W(k)-W(NT)\mid X\in G_{j}]$, $j, k \in \{T, NT, O\}$. By the randomization of arms (unconfoundedness) in our RCT data; the following holds 
\begin{equation}
E[W(k)-W(NT)\mid X\in G_{j}]=E[W \mid D = k, X \in G_{j}]-E[W\mid D = NT, X \in G_{j}], \label{eq:avg_W_given_G}
\end{equation}
The conditional expectations in the right-hand side can be estimated directly from our RCT data. For arm $j$, $E[W(T)-W(NT)\mid X\in G_{j}]$ corresponds to the (conditional) average treatment effect of the energy saving program, so we denote this quantity $\mathrm{ATE}$. On the other hand, $E[W(O)-W(NT) \mid X\in G_{j}]$ corresponds to the (conditional) intention to treat effect of the energy saving program, so we denote this quantity by $\mathrm{ITT}$. Note that when $k = NT$, $E[W(k)-W(NT)\mid X\in G_{j}] = 0$ always holds. In addition, we can identify the take-up rate when households were assigned to the opt-in arm,
\begin{equation}
P(Z(O)=T\mid X\in G_{j})=P(Z = T \mid D=O, X\in G_{j}). \label{eq:take-up rate}
\end{equation}
Assuming the exclusion restriction holds in the sense that a household's responses to the treatment is not causally affected by who makes the treatment choice, the planner or the households themselves, we can estimate the counterfactual average welfare effect for a households who is assigned to arm $j$ by policy $G$ and would have chosen $T$ if  assigned to arm $O$, as shown below. Identification the counterfactual welfare effect conditional on the assignment of arm hinges on this exclusion restriction, but identification of the optimal assignment policy and statistical properties of the estimated policy $\hat{G}^{mix}$ by EWM are unaffected no matter whether the exclusion restriction holds or not.

Under the exclusion restriction
\begin{equation}
    W(O) = W(T)\cdot 1\{Z(O)=T\} + W(NT)\cdot 1\{Z(O)=NT\}.
    \label{eq:decompose-W_O-under-ER}
\end{equation}
Decomposing $E[W(O)-W(NT)\mid X\in G_{j'}]$ using the law of iterated expectations and substituting \eqref{eq:decompose-W_O-under-ER} for $W(O)$ gives us 
\begin{multline*}
    E[W(O)-W(NT)\mid X\in G_{j}]
    =E[W(T)-W(NT)\mid Z(O)=T,X\in G_{j}]P(Z(O)=T\mid X\in G_{j}),
\end{multline*}
Plug in the quantities identified in \eqref{eq:avg_W_given_G} and \eqref{eq:take-up rate} to obtain
\begin{multline}
    E[W(T)-W(NT)\mid Z(O)=T,X\in G_{j}]\\
    =\frac{E[W \mid D = O, X \in G_{j}]-E[W\mid D = NT, X \in G_{j}]}{P(Z =T \mid D=O, X\in G_{j})}.
    \label{eq:late-of-complier}
\end{multline}
This identification result is analogous to the identification of the local average treatment effect (LATE) in \citep{Imbens1994}. Specifically, if we view an indicator for being assigned to arm $NT$ or $O$ as a binary instrumental variable and run two-stage least squares (2SLS) on the subsample of $\{i : D_i \in \{NT, O \}, X_i \in G_j \}$, treating the actual treatment take-up as an endogenous variable, we obtain (\ref{eq:late-of-complier}) as the 2SLS estimand. As $Z(O) = T$ corresponds to the subgroup of households who would choose to participate in the program if offered the autonomous choice, we refer to this quantity as \textit{LATE for takers}. LATE for takers relies on the comparison of households assigned to arms $O$ and $NT$ in the RCT data. A similar comparison of households assigned to arms $O$ and $T$ identifies the counterfactual average effect for households who are assigned to arm $j$ by policy $G$ and would choose $NT$ under arm $O$. Noting 
\begin{multline*}
E[W(T)-W(NT)\mid X\in G_{j}]\\
=E[W(T)-W(NT)\mid Z(O)=T,X\in G_{j}]P(Z(O)=T\mid X\in G_{j})\\
+E[W(T)-W(NT)\mid Z(O)=NT,X\in G_{j}]P(Z(O)=NT\mid X\in G_{j}),
\end{multline*}
and plugging in the quantities identified in \eqref{eq:avg_W_given_G}, \eqref{eq:take-up rate}, and \eqref{eq:late-of-complier}, we obtain 
\begin{multline}
    E[W(T)-W(NT)\mid Z(O)=NT,X\in G_{j}]\\
    =\frac{E[W\mid D=T, X\in G_{j}] - E[W\mid D=O, X\in G_{j}]}{1-P(Z=T\mid D=O,X\in G_{j})}.
    \label{eq:late-of-non-complier}
\end{multline}
Analogous to LATE for takers, this quantity corresponds to the estimand from a 2SLS regression using the subsample of $\{i : D_i \in \{T, O \}, X_i \in G_j \}$, where an indicator for being assigned to arm $T$ or $O$ is used as a binary instrumental variable for the actual treatment take-up. $Z(O) = NT$ corresponds to the subgroup households who would choose not to participate in the program if assigned to the autonomous choice arm, so we refer to this quantity as $\mathrm{LATE}$ for non-takers.\footnote{Given that the identification of the LATEs for takers and non-takers is analogous to the identification of LATE (for compliers) shown in \cite{Imbens1994}, we can derive a necessary testable implication for the exclusion restriction of household's responses to being assigned to arm $O$ following the testable implications of the LATE identifying assumptions obtained in \cite{balkepearl} and \cite{Heckman_2005}. Following the test approach of \cite{Kitagawa2015}, we plot the subdensities of the outcome distributions with our data and do not find a notable violation of the testable implications for the exclusion restriction.}

Panel A of Table~\ref{tab:mechanism} presents estimation results. Each column corresponds to a subgroup indexed by their assigned arm under the estimated optimal policy $\hat{G}^{mix}$. The first row shows estimated take-up rates if each were group were assigned to autonomous. These are based on the empirical analogue of equation \eqref{eq:take-up rate}. We observe that variance of the take-up rate across arms is low ($p$-value is 0.48).

\begin{center}
[Table \ref{tab:mechanism} about here]
\end{center}

The second and third rows show LATEs for takers and non-takers as estimated by equations \eqref{eq:late-of-complier} and \eqref{eq:late-of-non-complier}. Unlike the take-up rates, there is large variation in the LATE estimates between the takers and non-takers, and across subgroups. 

There are three main findings. First, among households assigned to arm $T$ by $\hat{G}^{mix}$, LATE for takers is 328.4 JPY and not significantly different from zero, whereas LATE for non-takers is 686.9 JPY and significantly different from zero. That is, non-takers in this subgroup positively contribute to welfare if they are exposed to the treatment, even though they would opt-out if given the choice. From the planner's point of view, it is preferable to paternalistically assign these households to $T$ instead of letting them choose. Second, for households assigned to arm $O$ under $\hat{G}^{mix}$, the estimated LATEs for takers and non-takers are 1369.6 JPY and -678.0 JPY respectively, and both are significantly different from zero. These estimates are consistent with households who would autonomously choose the treatment that is superior in terms of social welfare being assigned to $O$. The estimated policy $\hat{G}^{mix}$ assigns them to arm $O$ in order to benefit from this welfare gain. Finally, both estimated LATEs for households assigned to arm $NT$ are negative, and the LATE for takers is significantly different from zero. That is, among this subgroup, negative-selection of takers reduces social welfare and, for the planner, it is optimal to assign them to $NT$.

Panel B of Table~\ref{tab:mechanism} presents ATE and ITT as estimated by  \eqref{eq:avg_W_given_G}. Consistent with the heterogeneity of LATEs in Panel A, there is considerable variation in the ATEs and ITTs. For the subgroup of households assigned to arm $T$, the ATE and ITT estimates are 550.1 JPY and 125.3 JPY, and the difference between the two is significantly different ($p$-value 0.003). Consistent with their assignment under $\hat{G}^{mix}$, the welfare maximizing arm for this subgroup is $T$ (Recall that ATE and ITT correspond to the average welfare gains of arm $T$ and $O$). For households assigned to arm $O$, the ATE and ITT are 109.1 JPY and 526.5, and the difference between these estimates is significantly different from zero ($p$-value 0.002). In terms of social welfare, the optimal arm for this subgroup is $O$, implying that for these households autonomous choice dominates paternalistic assignment, and $\hat{G}^{mix}$ assigns them accordingly. For households recommended arm $NT$, the ATE and ITT are both negative, and both effects are different from zero ($p$-value is 0.000). The optimal arm for them is $NT$. 
 
 Hence, the arm recommended by $\hat{G}^{mix}$ coincides with the arm that attains the highest welfare gain for every subgroup of $\{X \in \hat{G}_j^{mix} \}$, $j \in \{T, NT, O \}$. In other words, the estimated policy $\hat{G}^{mix}$ captures households' heterogeneous responses, and mixes paternalistic assignments and autonomous choice to maximize social welfare.

\section{Conclusion}

We build an algorithm that mixes paternalistic and autonomous approaches to find an optimal targeting policy that maximizes policy impact. We apply this algorithm to a randomized field experiment involving an energy saving program. We show that optimally mixing paternalistic assignment and autonomous choice significantly improves the social welfare gain from the policy. To understand the mechanism behind our findings, we develop a method that exploits the random variation generated by our field experiment to estimate treatment effects. These estimates show that our algorithm optimally allocates individuals to be treated, untreated, or given an opt-in option based on the relative merits of paternalistic assignments and autonomous choice for their individual type.

\clearpage
\begin{spacing}{1.2}
\bibliographystyle{ecta}
\bibliography{ref,A_Bibdesk}

\begin{thebibliography}{45}
\newcommand{\enquote}[1]{``#1''}
\expandafter\ifx\csname natexlab\endcsname\relax\def\natexlab#1{#1}\fi

\bibitem[\protect\citeauthoryear{Alatas, Purnamasari, Wai-Poi, Banerjee, Olken,
  and Hanna}{Alatas et~al.}{2016}]{alatas_2016}
\textsc{Alatas, V., R.~Purnamasari, M.~Wai-Poi, A.~Banerjee, B.~A. Olken, and
  R.~Hanna} (2016): \enquote{Self-targeting: Evidence from a field experiment
  in Indonesia,} \emph{Journal of Political Economy}, 124, 371--427.

\bibitem[\protect\citeauthoryear{Andrews, Kitagawa, and McCloskey}{Andrews
  et~al.}{2019}]{Andrews_etal_2019}
\textsc{Andrews, I.~S., T.~Kitagawa, and A.~McCloskey} (2019):
  \enquote{Inference on Winners,} \emph{NBER working paper}.

\bibitem[\protect\citeauthoryear{Athey and Wager}{Athey and Wager}{2021}]{AW20}
\textsc{Athey, S. and S.~Wager} (2021): \enquote{Efficient policy learning with
  observational data,} \emph{Econometrica}, 89, 133--161.

\bibitem[\protect\citeauthoryear{Balke and Pearl}{Balke and
  Pearl}{1997}]{balkepearl}
\textsc{Balke, A. and J.~Pearl} (1997): \enquote{Bounds on treatment effects
  from studies with imperfect compliance,} \emph{Journal of the American
  Statistical Association}, 92, 1171--1176.

\bibitem[\protect\citeauthoryear{Bhattacharya}{Bhattacharya}{2013}]{Bhattacharya_2013}
\textsc{Bhattacharya, D.} (2013): \enquote{Evaluating Treatment Protocols by
  Combining Experimental and Observational Data,} \emph{Journal of
  Econometrics}, 173, 160--174.

\bibitem[\protect\citeauthoryear{Breiman, Friedman, Olshen, and Stone}{Breiman
  et~al.}{2017}]{breiman_et_al2017}
\textsc{Breiman, L., J.~H. Friedman, R.~A. Olshen, and C.~J. Stone} (2017):
  \emph{Classification and regression trees}, Routledge.

\bibitem[\protect\citeauthoryear{Burlig, Knittel, Rapson, Reguant, and
  Wolfram}{Burlig et~al.}{2020}]{burlig2020machine}
\textsc{Burlig, F., C.~Knittel, D.~Rapson, M.~Reguant, and C.~Wolfram} (2020):
  \enquote{Machine learning from schools about energy efficiency,}
  \emph{Journal of the Association of Environmental and Resource Economists},
  7, 1181--1217.

\bibitem[\protect\citeauthoryear{Cagala, Glogowsky, Rincke, and
  Strittmatter}{Cagala et~al.}{2021}]{cagala_2021}
\textsc{Cagala, T., U.~Glogowsky, J.~Rincke, and A.~Strittmatter} (2021):
  \enquote{Optimal Targeting in Fundraising: A Causal Machine-Learning
  Approach,} \emph{arXiv preprint arXiv:2103.10251}.

\bibitem[\protect\citeauthoryear{Chamberlain}{Chamberlain}{2011}]{Chamberlain2011}
\textsc{Chamberlain, G.} (2011): \enquote{Bayesian aspects of treatment
  choice,} in \emph{The Oxford Handbook of Bayesian Econometrics}, ed. by
  J.~Geweke, G.~Koop, and H.~van Dijk, Oxford University Press, 11--39.

\bibitem[\protect\citeauthoryear{Christensen, Francisco, Myers, Shao, and
  Souza}{Christensen et~al.}{2021}]{christensen_2021}
\textsc{Christensen, P., P.~Francisco, E.~Myers, H.~Shao, and M.~Souza} (2021):
  \enquote{Energy Efficiency Can Deliver for Climate Policy: Evidence from
  Machine Learning-Based Targeting,} .

\bibitem[\protect\citeauthoryear{Dehejia}{Dehejia}{2005}]{Dehejia2005}
\textsc{Dehejia} (2005): \enquote{Program evaluation as a decision problem,}
  \emph{Journal of Econometrics}, 125, 141--173.

\bibitem[\protect\citeauthoryear{Deshpande and Li}{Deshpande and
  Li}{2019}]{Deshpande_2019}
\textsc{Deshpande, M. and Y.~Li} (2019): \enquote{Who {Is} {Screened} {Out}?
  {Application} {Costs} and the {Targeting} of {Disability} {Programs},}
  \emph{American Economic Journal: Economic Policy}, 11, 213--248.

\bibitem[\protect\citeauthoryear{Dynarski, Libassi, Michelmore, and
  Owen}{Dynarski et~al.}{2018}]{dynarski_2018}
\textsc{Dynarski, S., C.~Libassi, K.~Michelmore, and S.~Owen} (2018):
  \enquote{Closing the gap: The effect of a targeted, tuition-free promise on
  college choices of high-achieving, low-income students,} Tech. rep., National
  Bureau of Economic Research.

\bibitem[\protect\citeauthoryear{Finkelstein and Notowidigdo}{Finkelstein and
  Notowidigdo}{2019}]{Finkelstein_2018}
\textsc{Finkelstein, A. and M.~J. Notowidigdo} (2019): \enquote{Take-up and
  {Targeting}: {Experimental} {Evidence} from {SNAP},} \emph{Quarterly Journal
  of Economics}, 134, 1505--1556.

\bibitem[\protect\citeauthoryear{Fowlie, Wolfram, Baylis, Spurlock, Todd-Blick,
  and Cappers}{Fowlie et~al.}{2021}]{Fowlie_et_al_2021}
\textsc{Fowlie, M., C.~Wolfram, P.~Baylis, C.~A. Spurlock, A.~Todd-Blick, and
  P.~Cappers} (2021): \enquote{Default Effects And Follow-On Behaviour:
  Evidence From An Electricity Pricing Program,} \emph{The Review of Economic
  Studies}, 88, 2886--2934.

\bibitem[\protect\citeauthoryear{Friedberg, Tibshirani, Athey, and
  Wager}{Friedberg et~al.}{2021}]{Friedberg2021}
\textsc{Friedberg, R., J.~Tibshirani, S.~Athey, and S.~Wager} (2021):
  \enquote{{Local Linear Forests},} \emph{Journal of Computational and
  Graphical Statistics}, 30, 503--517.

\bibitem[\protect\citeauthoryear{Gerarden and Yang}{Gerarden and
  Yang}{2021}]{gerarden_2021}
\textsc{Gerarden, T.~D. and M.~Yang} (2021): \enquote{Using Targeting to
  Optimize Program Design: Evidence from an Energy Conservation Experiment,}
  Tech. rep., Working Paper.

\bibitem[\protect\citeauthoryear{Heckman}{Heckman}{2010}]{Heckman_2010}
\textsc{Heckman, J.~J.} (2010): \enquote{Building {Bridges} between
  {Structural} and {Program} {Evaluation} {Approaches} to {Evaluating}
  {Policy},} \emph{Journal of Economic Literature}, 48, 356--398.

\bibitem[\protect\citeauthoryear{Heckman and Vytlacil}{Heckman and
  Vytlacil}{2005}]{Heckman_2005}
\textsc{Heckman, J.~J. and E.~Vytlacil} (2005): \enquote{Structural Equations,
  Treatment Effects, and Econometric Policy Evaluation,} \emph{Econometrica},
  73, 669--738.

\bibitem[\protect\citeauthoryear{Hirano and Porter}{Hirano and
  Porter}{2009}]{HiranoPorter2009}
\textsc{Hirano, K. and J.~R. Porter} (2009): \enquote{Asymptotics for
  statistical treatment rules,} \emph{Econometrica}, 77, 1683--1701.

\bibitem[\protect\citeauthoryear{Ida and Ushifusa}{Ida and
  Ushifusa}{2017}]{Ida2017}
\textsc{Ida, T. and Y.~Ushifusa} (2017): \enquote{{Cost-Benefit Analysis of
  Price-based Residential Demand Response},} \emph{Proceedings of the Japan
  Joint Automatic Control Conference}, 60, 304--307.

\bibitem[\protect\citeauthoryear{Imbens and Angrist}{Imbens and
  Angrist}{1994}]{Imbens1994}
\textsc{Imbens, G.~W. and J.~D. Angrist} (1994): \enquote{{Identification and
  Estimation of Local Average Treatment Effects},} \emph{Econometrica}, 62,
  467--475.

\bibitem[\protect\citeauthoryear{Ito, Ida, and Tanaka}{Ito
  et~al.}{2021}]{Ito_2019}
\textsc{Ito, K., T.~Ida, and M.~Tanaka} (2021): \enquote{Selection on Welfare
  Gains: Experimental Evidence from Electricity Plan Choice,} \emph{{NBER}
  Working Paper}.

\bibitem[\protect\citeauthoryear{Janevic, Janz, Dodge, Lin, Pan, Sinco, and
  Clark}{Janevic et~al.}{2003}]{Janevic_2003}
\textsc{Janevic, M.~R., N.~K. Janz, J.~A. Dodge, X.~Lin, W.~Pan, B.~R. Sinco,
  and N.~M. Clark} (2003): \enquote{The role of choice in health education
  intervention trials: a review and case study,} \emph{Social Science and
  Medicine}, 56, 1581--1594.

\bibitem[\protect\citeauthoryear{Johnson, Levine, and Toffel}{Johnson
  et~al.}{2020}]{johnson_2020}
\textsc{Johnson, M.~S., D.~I. Levine, and M.~W. Toffel} (2020):
  \enquote{Improving regulatory effectiveness through better targeting:
  Evidence from OSHA,} \emph{Harvard Business School Technology \& Operations
  Mgt. Unit Working Paper}.

\bibitem[\protect\citeauthoryear{Kitagawa}{Kitagawa}{2015}]{Kitagawa2015}
\textsc{Kitagawa, T.} (2015): \enquote{A Test for Instrument Validity,}
  \emph{Econometrica}, 83, 2043--2063.

\bibitem[\protect\citeauthoryear{Kitagawa and Tetenov}{Kitagawa and
  Tetenov}{2018}]{KT18}
\textsc{Kitagawa, T. and A.~Tetenov} (2018): \enquote{Who should be treated?
  Empirical welfare maximization methods for treatment choice,}
  \emph{Econometrica}, 86, 591--616.

\bibitem[\protect\citeauthoryear{Knittel and Stolper}{Knittel and
  Stolper}{2019}]{knittel2019using}
\textsc{Knittel, C.~R. and S.~Stolper} (2019): \enquote{Using machine learning
  to target treatment: The case of household energy use,} Tech. rep., National
  Bureau of Economic Research.

\bibitem[\protect\citeauthoryear{Knox, Yamamoto, Baum, and Berinsky}{Knox
  et~al.}{2019}]{Knox_etal_2019}
\textsc{Knox, D., T.~Yamamoto, M.~A. Baum, and A.~J. Berinsky} (2019):
  \enquote{Design, Identification, and Sensitivity Analysis for Patient
  Preference Trials,} \emph{Journal of the American Statistical Association},
  114, 1532--1546.

\bibitem[\protect\citeauthoryear{Lieber and Lockwood}{Lieber and
  Lockwood}{2019}]{lieber_2019}
\textsc{Lieber, E.~M. and L.~M. Lockwood} (2019): \enquote{Targeting with
  in-kind transfers: Evidence from Medicaid home care,} \emph{American Economic
  Review}, 109, 1461--85.

\bibitem[\protect\citeauthoryear{Long, Little, and Lin}{Long
  et~al.}{2008}]{Long_etal_2008}
\textsc{Long, Q., R.~Little, and X.~Lin} (2008): \enquote{Causal Inference in
  Hybrid Intervention Trials Involving Treatment Choice,} \emph{Journal of the
  American Statistical Association}, 103, 474--484.

\bibitem[\protect\citeauthoryear{Manski}{Manski}{2013}]{ManskiBook13}
\textsc{Manski, C.} (2013): \emph{Public Policy in an Uncertain World},
  Cambridge, MA: Harvard University Press.

\bibitem[\protect\citeauthoryear{Manski}{Manski}{2004}]{manski2004statistical}
\textsc{Manski, C.~F.} (2004): \enquote{Statistical treatment rules for
  heterogeneous populations,} \emph{Econometrica}, 72, 1221--1246.

\bibitem[\protect\citeauthoryear{Murakami, Shimada, Ushifusa, and Ida}{Murakami
  et~al.}{2020}]{murakami_2020}
\textsc{Murakami, K., H.~Shimada, Y.~Ushifusa, and T.~Ida} (2020):
  \enquote{Heterogeneous Treatment Effects of Nudge and Rebate: Causal Machine
  Learning in a Field Experiment on Electricity Conservation,} \emph{Kyoto
  University, Graduate School of Economics Discussion Paper Series, Discussion
  Paper No. E-20-003.}

\bibitem[\protect\citeauthoryear{Potter, Jimenez, and George}{Potter
  et~al.}{2014}]{Potter_et_al_2014}
\textsc{Potter, J., L.~Jimenez, and S.~George} (2014): \enquote{SmartPricing
  Options final evaluation: The final report on pilot design, implementation,
  and evaluation of the Sacramento Municipal Utility District’s Consumer
  Behavior Study,} \emph{Sacramento Municipal Utility District}.

\bibitem[\protect\citeauthoryear{R\"{u}cker}{R\"{u}cker}{1989}]{Rucker_1989}
\textsc{R\"{u}cker, G.} (1989): \enquote{A Two-Stage Trial Design for Testing
  Treatment, Self-Selection, and Treatment Preference Effects,}
  \emph{Statistics in Medicine}, 8, 477--485.

\bibitem[\protect\citeauthoryear{Stoye}{Stoye}{2009}]{stoye2009minimax}
\textsc{Stoye, J.} (2009): \enquote{Minimax regret treatment choice with finite
  samples,} \emph{Journal of Econometrics}, 151, 70--81.

\bibitem[\protect\citeauthoryear{Stoye}{Stoye}{2012}]{stoye2012minimax}
---\hspace{-.1pt}---\hspace{-.1pt}--- (2012): \enquote{Minimax regret treatment
  choice with covariates or with limited validity of experiments,}
  \emph{Journal of Econometrics}, 166, 138--156.

\bibitem[\protect\citeauthoryear{Tetenov}{Tetenov}{2012}]{tetenov2012statistical}
\textsc{Tetenov, A.} (2012): \enquote{Statistical treatment choice based on
  asymmetric minimax regret criteria,} \emph{Journal of Econometrics}, 166,
  157--165.

\bibitem[\protect\citeauthoryear{Unrath}{Unrath}{2021}]{unrath_2021}
\textsc{Unrath, M.} (2021): \enquote{Targeting, Screening, and Retention:
  Evidence from California’s Food Stamps Program,} .

\bibitem[\protect\citeauthoryear{Wager and Athey}{Wager and
  Athey}{2018}]{Wager2018}
\textsc{Wager, S. and S.~Athey} (2018): \enquote{{Estimation and Inference of
  Heterogeneous Treatment Effects using Random Forests},} \emph{Journal of the
  American Statistical Association}, 113, 1228--1242.

\bibitem[\protect\citeauthoryear{Waldinger}{Waldinger}{2021}]{waldinger_2021}
\textsc{Waldinger, D.} (2021): \enquote{Targeting in-kind transfers through
  market design: A revealed preference analysis of public housing allocation,}
  \emph{American Economic Review}, 111, 2660--96.

\bibitem[\protect\citeauthoryear{Wang and Tang}{Wang and
  Tang}{2018}]{Wang_2018}
\textsc{Wang, X. and W.~Tang} (2018): \enquote{To Overconsume or Underconsume:
  Baseline Manipulation in Demand Response Programs,} \emph{North American
  Power Symposium}.

\bibitem[\protect\citeauthoryear{Wolak}{Wolak}{2007}]{Wolak_2007}
\textsc{Wolak, F.~A.} (2007): \enquote{Residential customer response to
  real-time pricing: The anaheim critical peak pricing experiment,} \emph{UC
  Berkeley: Center for the Study of Energy Markets}.

\bibitem[\protect\citeauthoryear{Zhou, Athey, and Wager}{Zhou
  et~al.}{2018}]{zhou_et_al_2018}
\textsc{Zhou, Z., S.~Athey, and S.~Wager} (2018): \enquote{Offline multi-action
  policy learning: Generalization and optimization,} \emph{arXiv preprint
  arXiv:1810.04778}.

\end{thebibliography}
\end{spacing}

\clearpage
\section*{Figures}

\begin{figure}[H]
    \centering
    \caption{Example of $G^{\ast}$}
    \label{fig:example_optimal_policy}
\begin{tikzpicture}
\fill[red!50] (3.5,6)  -- (3.5,3.5/2) -- (9,1.5)  -- (9,6) -- cycle;
\fill[yellow!50] (0,6) -- (3.5,4.25) -- (3.5,6) -- cycle;
\fill[yellow!50] (3.5,0) -- (9,0) -- (9,1.5) -- (3.5,4.25) -- cycle;
\fill[blue!50] (0,0) -- (0,6) -- (3.5,4.25) -- (3.5,0) -- cycle;
\draw[dashed, pattern={Lines[angle=45,distance=8pt]}, pattern color=gray!150]
(0,6) -- (9,1.5) -- (9,6) -- cycle;
\draw[dashed, pattern={Lines[angle=-45,distance=8pt]}, pattern color=gray!150]
(0,0) -- (9,4.5) -- (9,0) -- cycle;
\draw[dashed, pattern={Lines[angle=90,distance=8pt]}, pattern color=gray!150]
(3.5,0) -- (9,0) -- (9,6) -- (3.5,6) -- cycle;
\draw (0,0) -- (9,0) -- (9,6) -- (0,6) -- (0,0);
\node[anchor=north] at (8.75,0) {$X_1$};
\node[anchor=east] at (0,5.75) {$X_2$};
\matrix[draw,below right] at (current bounding box.north east) {
 \node [rednode,label=right:$G_{T}^{\ast}$] {}; \\
 \node [bluenode,label=right:$G_{NT}^{\ast}$] {}; \\
 \node [yellownode,label=right:$G_{O}^{\ast}$] {}; \\
  \node [dotnode,label=right:$ \{x \mid CATE(x) \geq 0\}$] {}; \\
  \node [slashnode,label=right:$ \{x \mid CATE_{T}(x) \geq 0\}$] {}; \\
  \node [backslashnode,label=right:$\{x \mid CATE_{NT}(x) \geq 0\}$]{}; \\
};
\end{tikzpicture}
\begin{tablenotes}\footnotesize
\item[] Notes: $CATE(x) := E[W(T)-W(NT) \mid x]$, $CATE_{T}(x) := E[W(T)-W(NT) \mid Z(O)=T,x]$, and $CATE_{NT}(x) := E[W(T)-W(NT) \mid Z(O)=NT,x]$.
\end{tablenotes} 
\end{figure}

\clearpage
\begin{figure}[H]
\caption{Examples of depth-1 and -2 decision trees.} \label{fig:example_decision_trees}
\begin{tabular}{cc}
\begin{minipage}[t]{0.5\hsize}
\begin{center}
{\small (a-1) Depth 1 decision tree}

\bigskip
\bigskip
\smallskip

\begin{tikzpicture}[scale=1.7]
 
\node [draw,rounded corners] {\Large $X_{1} \geq a_{1}$}[sibling distance = 2cm]
    child {node [draw,rounded corners] {\Large T} edge from parent node [left] {{True}}}
    child {node [draw,rounded corners] {\Large NT} edge from parent node [right]  {False}};
 
\end{tikzpicture}
 
\end{center}
\end{minipage}
\begin{minipage}[t]{0.5\hsize}
\begin{center}
{\small (a-2) Partition of $\MX$ by the depth 1 decision tree in (a-1)}
\medskip

\begin{tikzpicture}
\draw (3.3,0) -- (3.3,4);
\draw (0,0) -- (6,0) -- (6,4) -- (0,4) -- (0,0);
\node[anchor=north] at (5.9,0) {$X_1$};
\node[anchor=east] at (0,3.9) {$X_2$};
\node[anchor=north] at (3.3,0) {$a_{1}$};
\node at (4.65,2) {\LARGE $G_{T}$};
\node at (1.65,2) {\LARGE $G_{NT}$};
\end{tikzpicture}

\end{center}
\end{minipage}
\end{tabular}
\begin{tabular}{cc}
\begin{minipage}[t]{0.5\hsize}
\begin{center}
{\small (b-1) Depth 2 decision tree}

\bigskip
\smallskip

\begin{tikzpicture}
[   scale = 1.1,
    level 1/.style = {sibling distance = 3.5cm},
    level 2/.style = {sibling distance = 2cm}
]
 
\node [draw,rounded corners] {\large  $X_{1} \geq a_{1}$}
    child {node [draw,rounded corners] {\large $X_{2} \geq a_{2}$} 
    child {node [draw,rounded corners][draw,rounded corners]  {\large T} edge from parent node [left] {True}}
    child {node [draw,rounded corners] {\large O} edge from parent node [right] { False}}
    edge from parent node [left] {True}} 
    child {node [draw,rounded corners] {\large $X_{1} \geq a_{3}$}
    child {node [draw,rounded corners] {\large NT} edge from parent node [left] { True}}
    child {node [draw,rounded corners] {\large O} edge from parent node [right] { False}}
    edge from parent node [right] {False}
    };
 
\end{tikzpicture}
\end{center}
\end{minipage}
\begin{minipage}[t]{0.5\hsize}
\begin{center}
{\small (b-2) Partition of $\MX$ by the depth 2 decision tree in (b-1)}
\medskip

\begin{tikzpicture}
\draw (3.3,0) -- (3.3,4);
\draw (1.65,0) -- (1.65,4);
\draw (3.3,2) -- (6,2);
\draw (0,0) -- (6,0) -- (6,4) -- (0,4) -- (0,0);
\node[anchor=north] at (5.9,0) {$X_1$};
\node[anchor=east] at (0,3.9) {$X_2$};
\node[anchor=north] at (3.3,0) {$a_{1}$};
\node[anchor=east] at (0,2) {$a_{2}$};
\node[anchor=north] at (1.65,0) {$a_{3}$};
\node at (1.65/2,2) {\LARGE $G_{O}$};
\node at (4.95/2,2) {\LARGE $G_{NT}$};
\node at (9.3/2,3) {\LARGE $G_{T}$};
\node at (9.3/2,1) {\LARGE $G_{O}$};
\end{tikzpicture}

\end{center}
\end{minipage}
\end{tabular}
\begin{tablenotes}\footnotesize
\item[] Notes: Panels (a-2) and (b-2) correspond to the partitions of $\MX$ induced by the decision trees in Panels (a-1) and (a-2), respectively, where we assume $\MX$ is a two-dimensional space and $a_1 > a_3$.
\end{tablenotes} \label{fig:empirics_CI of T}
\end{figure}

\clearpage

\clearpage

\clearpage 

\clearpage 

\clearpage
\section*{Tables}

\begin{center}
\begin{table}[H]
\caption{\label{tab:summary}Summary Statistics}
\begin{centering}
\begin{tabular*}{\textwidth}{l @{\extracolsep{\fill}} ccccccc}
\toprule
 & & & & \multicolumn{4}{c}{p-value: $H_0$: difference in means = 0}   \\
 \cmidrule(l{3pt}r{3pt}){5-8} 
 & NT & T & O & NT vs. T & NT vs. O & T vs. O & ALL \\
\midrule
Peak hour usage (kWh)  & 0.192 & 0.190 & 0.189 & 0.609 & 0.633 & 0.961 & 0.838 \\
        & (0.004) & (0.004) & (0.005) &  &  &  &  \\
        \addlinespace 
Pre-peak hour usage (kWh) & 0.179 & 0.176 & 0.180 & 0.442 & 0.854 & 0.417 & 0.646 \\
        & (0.003) & (0.004) & (0.005) &  &  &  &  \\
        \addlinespace 
Post-peak hour usage (kWh)  & 0.299 & 0.297 & 0.293 & 0.756 & 0.426 & 0.588 & 0.725 \\
        & (0.004) & (0.004) & (0.006) &  &  &  &  \\
        \addlinespace 
Number of people at home & 2.481 & 2.444 & 2.470 & 0.406 & 0.846 & 0.626 & 0.703 \\
        & (0.031) & (0.032) & (0.045) &  &  &  &  \\
        \addlinespace 
Interest in energy conservation & 3.450 & 3.464 & 3.488 & 0.647 & 0.299 & 0.517 & 0.584 \\
        (1-5 Likert scale) & (0.022) & (0.022) & (0.029) &  &  &  &  \\
        \addlinespace 
Household income & 645.095 & 613.409 & 636.641 & 0.022 & 0.622 & 0.154 & 0.066 \\
        (10,000JPY) & (10.056) & (9.384) & (13.778) &  &  &  &  \\
        \addlinespace 
Number of customers & 1,577 & 1,486 & 807 & & & & \\
        \bottomrule
\end{tabular*}

\par\end{centering}
\vspace{.1cm}
{\small{}Notes: The first three columns present the sample mean and standard errors for the pre-experiment consumption data and demographic variables. The first column contains values for the group assigned to no-treatment ($NT$), the second contains values for the group assigned to treatment ($T$), and the third contains values for the group given the option to receive treatment ($O$). We do not observe any significant difference in the first five variables between the groups. This indicates that random assignment has successfully balanced these five observables. For household income, we do not observe any significant difference between the $NT$ and $O$ groups, nor the $T$ and $O$ groups, whereas we do observe a significant difference between the $NT$ and $T$ groups. We confirmed that the treatment effects did not vary substantially when estimating the effects while controlling for demographic variables including household income.}{\small\par}
\end{table}
\newpage{}
\par\end{center}

\begin{center}
\begin{table}[H]
\caption{\label{tab:ITT} Intention-to-Treat Estimates}
\begin{centering}
\begin{tabular*}{\textwidth}{l @{\extracolsep{\fill}} ccccc}
    \toprule
&  & \multicolumn{2}{c}{Peak hour usage} & \multicolumn{2}{c}{Peak hour usage} \\
&  & \multicolumn{2}{c}{$-$ Pre-peak hour usage} & \multicolumn{2}{c}{$-$ Post-peak hour usage} \\ 
&  & \multicolumn{2}{c}{(in pre-experiment)} & \multicolumn{2}{c}{(in pre-experiment)}  \\ 
    \cmidrule(l{3pt}r{3pt}){3-6} 
& All &  Low & High &  Low & High   \\ 
    \midrule
100\% Treatment & -0.097  & -0.108  & -0.079  & -0.089  & -0.094    \\
    & (0.021) & (0.028) & (0.031) & (0.030) & (0.028)   \\
    \addlinespace 
100\% Opt-in & -0.052  & -0.022  & -0.073  & -0.070  & -0.023   \\
    & (0.027) & (0.034) & (0.041) & (0.037) & (0.037)   \\
    \midrule
Number of customers & 3,870 & 1,935 & 1,935 & 1,937 & 1,933 \\
Number of observations & 1,176,480 & 588,240  & 588,240  & 589,152  & 587,328 \\
p-value (T = O) & 0.088  & 0.013  & 0.880  & 0.595  & 0.047 \\
Opt-in rate & 37.2\%  & 36.9\%  & 37.4\%  & 39.9\%  & 34.7\% \\
    \bottomrule
\end{tabular*}

\vspace{.5cm}

\begin{tabular*}{\textwidth}{l @{\extracolsep{\fill}} cccccc}
    \toprule
& \multicolumn{2}{c}{Number of people at home}  & \multicolumn{2}{c}{Interest in} & \multicolumn{2}{c}{Household income} \\
& & & \multicolumn{2}{c}{energy conservation} & \\
    \cmidrule(l{3pt}r{3pt}){2-7} 
&  Low & High &  Low & High &  Low & High  \\ 
    \midrule
100\% Treatment & -0.096  & -0.098  & -0.134  & -0.057  & -0.071  & -0.125  \\
    & (0.027) & (0.034) & (0.028) & (0.031) & (0.028) & (0.031)   \\
    \addlinespace 
100\% Opt-in & -0.022  & -0.094  & -0.036  & -0.072  & -0.036  & -0.060 \\
    & (0.034) & (0.042) & (0.035) & (0.040) & (0.038) & (0.037)   \\
    \midrule
Number of customers  & 2,245 & 1,625 & 1,967 & 1,903 & 2,036 & 1,834 \\
Number of observations & 682,480  & 494,000  & 597,968  & 578,512  & 618,944  & 557,536 \\
p-value (T = O) & 0.020  & 0.934  & 0.004  & 0.715  & 0.336  & 0.094    \\
Opt-in rate & 37.6\%  & 36.6\%  & 33.8\%  & 40.6\%  & 34.8\%  & 39.7\%  \\
    \bottomrule
\end{tabular*}

\par\end{centering}
\vspace{.1cm}
{\small{}Notes: This table shows the estimation results for equation (6) using the full-sample (the first column of the upper panel) or sub-samples (the remaining columns). The dependent variable is the log of household-level electricity consumption over a 30-minute interval . We include household fixed effects and time fixed effects for each 30-minute interval. The standard errors are clustered at the household level to adjust for serial correlation. In order to investigate the heterogeneity of the treatment effects, we divided the sample into five sets of two sub-groups. For five different variables, the first sub-group includes households who are below the median of this variable and the second includes those who are above the median. }{\small\par}
\end{table}
\newpage{}
\par\end{center}
\newpage

\begin{table}
\centering
\begin{threeparttable}
\caption{Welfare performance of the best paternalistic assignment}

\begin{tabular*}{\textwidth}{l @{\extracolsep{\fill}} cr}
\toprule
Policy & Est. Welfare Gain & 95 \% CI\\
\midrule
Uniform no-treatment (100\% $NT$) & $0.0$ & ---\\
Uniform treatment (100\% $T$) & $63.1$ & $( -97.2 , 223.4 )$\\
The purely autonomous policy (100\% $O$) & $140.9$ & $( -40.1 , 322.0 )$\\
The paternalistic assignment ($\hat{G}^{pat}$) & $228.5$ & $( 109.7 , 347.4 )$\\
\midrule
$\hat{G}^{pat}$ vs 100\% $T$ & $165.5$ & $( 58.3 , 272.7 )$\\
$\hat{G}^{pat}$ vs 100\% $O$ & $87.6$ & $( -88.9 , 264.1 )$\\
\bottomrule
\end{tabular*}

\label{tab:purely-paternalistic-approach}
\begin{tablenotes}[flushleft]
    \footnotesize\item[]Notes: This table shows the estimated welfare gain of the optimal paternalistic assignment along with three benchmark policies. The unit of these estimates is Japanese yen. 95\% confidence intervals are in parentheses. The first four rows show the welfare gain relative to the status quo (uniform no-treatment), while the remaining two rows show the welfare gain of paternalistic assignment relative to the other two benchmark policies.
\end{tablenotes}
\end{threeparttable}
\end{table}

\begin{table}
\centering
\begin{threeparttable}
\caption{Welfare performance of the best mixture of the paternalistic and autonomous approaches}

\begin{tabular*}{\textwidth}{l @{\extracolsep{\fill}} cr}
\toprule
Policy & Est. Welfare Gain & 95 \% CI\\
\midrule
Uniform no-treatment (100\% $NT$) & $0.0$ & ---\\
Uniform treatment (100\% $T$) & $63.1$ & $( -97.2 , 223.4 )$\\
The purely autonomous policy (100\% $O$) & $140.9$ & $( -40.1 , 322.0 )$\\
The paternalistic assignment ($\hat{G}^{pat}$) & $228.5$ & $( 109.7 , 347.4 )$\\
The mix of paternalism and autonomy ($\hat{G}^{mix}$) & $437.9$ & $( 283.0 , 592.8 )$\\
\midrule
\addlinespace
$\hat{G}^{mix}$ vs 100\% $T$ & $374.8$ & $( 238.8 , 510.8 )$\\
$\hat{G}^{mix}$ vs 100\% $O$ & $297.0$ & $( 165.9 , 428.0 )$\\
$\hat{G}^{mix}$ vs $\hat{G}^{pat}$ & $209.4$ & $( 75.2 , 343.5 )$\\
\bottomrule
\end{tabular*}

\label{tab:mixture-approach}
\begin{tablenotes}[flushleft]
    \footnotesize\item[]Notes: This table shows the estimated welfare gain of the best mixture of paternalistic assignment and autonomous choice, along with that of three benchmark policies and the best paternalistic assignment. The unit of these estimates is Japanese yen. 95\% confidence intervals are in parentheses. The first five rows show the welfare gain relative to the status quo (uniform no-treatment), while the remaining three rows show the welfare gain of the best mixture relative to the other two benchmark policies and the optimal paternalistic assignment.
\end{tablenotes}
\end{threeparttable}
\end{table}

\clearpage
\begin{table}
\centering
\begin{threeparttable}
\caption{Mechanism behind the algorithm}

\begin{tabular*}{\textwidth}{l @{\extracolsep{\fill}} ccc}
\toprule
\multicolumn{1}{c}{ } & \multicolumn{3}{c}{Recommended arm by $\hat{G}^{mix}$} \\
\cmidrule(l{3pt}r{3pt}){2-4}
$j$ & $NT$ & $T$ & $O$\\
\midrule
\addlinespace[0.3em]
\multicolumn{4}{l}{\textbf{A) Counterfactual analysis of opt-in policy}}\\
\hspace{1em}Opt-in rate $(X \in \hat{G}_{j}^{mix})$ & $43.7\%$ & $38.2\%$ & $38.4\%$\\
\hspace{1em} & $(36\%,52\%)$ & $(33\%,44\%)$ & $(33\%,43\%)$\\
\hspace{1em}$\mathrm{LATE}(Z(O)=T, X \in \hat{G}_{j}^{mix})$ & $-1629.3$ & $328.4$ & $1369.6$\\
\hspace{1em} & $(-2602.5,-656.2)$ & $(-462.4,1119.3)$ & $(649.8,2089.5)$\\
\hspace{1em}$\mathrm{LATE}(Z(O)=NT, X \in \hat{G}_{j}^{mix})$ & $-518.4$ & $686.9$ & $-678.0$\\
\hspace{1em} & $(-1268.2,231.3)$ & $(231.8,1142.0)$ & $(-1102.4,-253.7)$\\
\addlinespace[0.3em]
\hline
\multicolumn{4}{l}{\textbf{B) Counterfactual analysis of each policy}}\\
\hspace{1em}$\mathrm{ATE}(X \in \hat{G}_{j}^{mix})$ & $-1004.5$ & $550.1$ & $109.1$\\
\hspace{1em} & $(-1364.5,-644.4)$ & $(284.8,815.5)$ & $(-130.3,348.5)$\\
\hspace{1em}$\mathrm{ITT}(X \in \hat{G}_{j}^{mix})$ & $-712.8$ & $125.3$ & $526.5$\\
\hspace{1em} & $(-1117.6,-308.1)$ & $(-175.9,426.6)$ & $(258.5,794.5)$\\
\bottomrule
\end{tabular*}

\label{tab:mechanism}
\begin{tablenotes}[flushleft]
    \footnotesize\item[]Notes: This table shows the estimates of the opt-in rate, LATE for takers, LATE for non-takers, ATE, and ITT. Each column contains estimates for a group of households recommended a particular arm by the estimated policy $\hat{G}^{mix}$. The unit of all estimates except the opt-in rate is JPY. 95\% confidence intervals are in parentheses. These estimates are calculated based on the identification results \eqref{eq:avg_W_given_G}, \eqref{eq:take-up rate}, \eqref{eq:late-of-complier}, and \eqref{eq:late-of-non-complier}.
\end{tablenotes}
\end{threeparttable}
\end{table}

\clearpage
\appendix
\section{Appendix}

\subsection{The external validity of the experimental sample \label{sec:externalvalidity}}

We randomly sampled 2,070 customers from the target population who did not participate in this experiment, and conducted a similar survey to the one for the experimental sample. The purpose of this was to investigate the external validity of our experimental sample by comparing the mean for each variable between the control group from our experimental sample and this random sample. Columns 1 and 2 of Table Appendix 1 show summary statistics for the control group and random sample. Column 3 presents differences in means, with the standard errors of these differences in brackets. We observe larger means for four variables in the control group than in the random sample, and the differences are statistically significant. Our experimental sample has larger pre-experiment electricity usage per month, a larger number of people at home on weekdays, higher interest in energy conservation, and higher household income. This implies that our sample includes a larger number of customers who are willing and able to reduce their electricity consumption, which should be taken into consideration when discussing the generalizability of this study.


\vskip\baselineskip
\vskip\baselineskip

\begin{center}
\begin{table}[H]
\renewcommand{\tablename}{Table Appendix}{}
\setcounter{table}{0}
\caption{\label{tab:The external validity of the sample} The external validity of the experimental sample}
\begin{centering}
\resizebox{\textwidth}{!}{\begin{tabular}{lccc}
    \toprule
& \multicolumn{1}{c}{Experimental sample} & \multicolumn{1}{c}{Random sample} & \multicolumn{1}{c}{Difference} \\
& \multicolumn{1}{c}{in the control group} & \multicolumn{1}{c}{of population} & \multicolumn{1}{c}{between sample} \\
& & & \multicolumn{1}{c}{and population} \\
    \midrule
Monthly electricity usage in July (kWh) & 355.703 & 303.841 & 51.862 \\
& (205.267) & (176.677) & [6.336] \\
    \addlinespace 
Number of people at home & 2.481 & 2.313 & 0.168 \\
& (1.241) & (1.204) & [0.041] \\
    \addlinespace 
Interest in energy conservation & 3.450 & 3.322 & 0.128 \\
(1-5 Likert scale) & (0.854) & (0.969) & [0.031] \\
    \addlinespace 
Household income (10,000JPY) & 645.095 & 581.264 & 63.832 \\
& (399.342) & (383.782) & [13.055] \\
    \addlinespace
    \addlinespace 
Number of customers & 1,577 & 2,070 &  \\
    \bottomrule
\end{tabular}
}
\par\end{centering}
\vspace{.5cm}
\end{table}
\newpage{}
\par\end{center}

\subsection{Two-step Procedure}

We describe the heuristic two-step procedure to obtain a decision tree of depth 6. Our procedure begins by arbitrarily choosing two arms from the set of all arms. For exposition purpose, suppose these are $T$ and $NT$. Given this pair of chosen arms, we first search for the best decision tree of depth 3 that exactly maximizes the empirical welfare function with the available arms restricted to $T$ and $NT$. The resulting tree distributes the training sample among the leaf nodes, and in each leaf node either of $T$ or $NT$ is selected as the best arm. Next, for each leaf node, we again search for the best decision tree of depth 3, but the available arms are modified to the current best arm ($T$ or $NT$) and the excluded arm ($O$). Concatenating these decision trees gives us a decision tree of depth 6. Notice that this procedure is dependent on the choice of the first two arms, $(T,NT)$, $(NT,O)$, and $(T,O)$. Since our goal is to approximate the decision tree of depth 6 which attains the highest empirical welfare, we repeat this two step procedure for all possible starting pairs and choose the one with the highest welfare performance. In our analysis, the decision tree with first step options  $NT$ and $O$ is selected. Although there is no guarantee that our procedure well approximates a global optimizer, the welfare performance of a policy obtained in this manner constitutes a lower bound for that of a global optimizer. Therefore, while our decision tree is conservative in terms of welfare, and we do not regard this as a serious drawback.

\subsection{Artificial Test Data}

We describe in detail our method for generating artificial test data. To begin with, we describe the source of the bias in point estimates. This occurs due to noise in $Y_i$ and $Z_i$, which are random components in $W_i$. Specifically, the observed outcome $Y_i$ can be decomposed as the sum of an essential term and noise as follows:
\begin{multline*}
    Y_i = \underbrace{E[Y_i(NT)\mid X_i]\cdot 1\{D_i = NT\} + E[Y_i(T) \mid X_i] \cdot 1\{D_i = T\} + E[Y_i(O) \mid X_i]\cdot 1\{D_i = O\}}_{\text{essential term}}\\
    + \underbrace{\epsilon_i(NT)\cdot 1\{D_i = NT\} + \epsilon_i(T)\cdot 1\{D_i = T\} + \epsilon_i(O)\cdot 1\{D_i = O\}}_{\text{noise}},
\end{multline*}
where $\epsilon_i(j) := Y_i(j) - E[Y_i(j) \mid X_i]$ for each $j \in \{T,NT,O\}$. The observed choice $Z_i$ can be similarly decomposed. While only the first term is necessary for learning an optimal policy, a learning algorithm inevitably responds to the noise term and overfits the training sample at hand. When we evaluate the welfare performance of the estimated policy on the same training sample, the welfare estimate is biased upward because the policy fits the noise term as well. This implies that, if we replace the noise in the training sample with a second independent sample, we can eliminate the bias from the estimate of welfare performance.

Motivated by this, we generate test data $\{Y_i^{\text{test}}, Z_i^{\text{test}}, D_i, X_i\}_{i=1}^n$, where $Y_i^{\text{test}}$ denotes electricity consumption and $Z_i^{\text{test}}$ denotes treatment choice. For  $Y_i^{\text{test}}$ we use the following procedure to generate artificial data: For samples $i \in I_j := \{i : D_i = j\}$,
\begin{enumerate}
    \item Estimate $E[Y_i(j)\mid X_i]$ and calculate residuals $\widehat{\epsilon}_i = Y_i - \widehat{E}[Y_i(j)\mid X_i]$ for each $i \in I_j$.
    \item Sample $\{\widehat{\epsilon}_i^{\text{test}}\}_{i \in I_j}$ from $\{\widehat{\epsilon}_i\}_{i \in I_j}$ with replacement.
    \item Construct $Y_i^{\text{test}} = \widehat{E}[Y_i(j) \mid X_i] + \widehat{\epsilon}_i^{\text{test}}$ for each $i \in I_j$.
\end{enumerate}
Note that we implicitly assume homoskedasticity. In the first step, we estimate the conditional mean using a local linear forest \citep{Friedberg2021}. We generate $\{Z_i^{\text{test}}\}_{i=1}^n$ as follows: For samples $i \in I_{NT}$ and $i \in I_T$, we set $Z_i^{\text{test}} = NT$ and $Z_i^{\text{test}} = T$, respectively. For samples $i \in I_O$
\begin{enumerate}
    \item Estimate $P(Z_i(O) = T \mid X_i)$ for each $i \in I_O$.
    \item Sample $\{Z_i^{\text{test}}\}_{i \in I_O}$ such that $Z_i^{\text{test}} \sim \widehat{P}(Z_i(O) = T \mid X_i)$.
\end{enumerate}

\newpage

\end{document}